\documentclass[twocolumn]{aastex7}

\usepackage{lipsum}  
\usepackage{amsmath}
  
\usepackage{mathtools}
\usepackage{bm}       
\usepackage{mathrsfs} 
\usepackage{graphicx} 
\graphicspath{{fig/fig_pdf/}} 
\usepackage{booktabs} 
\usepackage{makecell} 
\usepackage{float} 

\newcommand{\erfc}{\operatorname{erfc}} 
\usepackage{hyperref} 
\begin{document}

\title{Exploring Joint Observation of the CSST Shear and clustering of astrophysical gravitational wave source measurements}

\correspondingauthor{Yan Gong}
\email{Email: gongyan@bao.ac.cn}

\author{Pengfei Su}
\affiliation{National Astronomical Observatories, Chinese Academy of Sciences, 20A Datun Road, Beijing 100101, China}
\affiliation{School of Astronomy and Space Sciences, University of Chinese Academy of Sciences(UCAS), \\Yuquan Road NO.19A Beijing 100049, China}
\email{supf@bao.ac.cn}

\author[orcid=0000-0003-0709-0101]{Yan Gong}
\affiliation{National Astronomical Observatories, Chinese Academy of Sciences, 20A Datun Road, Beijing 100101, China}
\affiliation{School of Astronomy and Space Sciences, University of Chinese Academy of Sciences(UCAS), \\Yuquan Road NO.19A Beijing 100049, China}
\affiliation{Science Center for Chinese Space Station Survey Telescope, National Astronomical Observatories, \\Chinese Academy of Science, 20A Datun Road, Beijing 100101, China}
\email{gongyan@bao.ac.cn}

\author{Qi Xiong}
\affiliation{National Astronomical Observatories, Chinese Academy of Sciences, 20A Datun Road, Beijing 100101, China}
\affiliation{School of Astronomy and Space Sciences, University of Chinese Academy of Sciences(UCAS), \\Yuquan Road NO.19A Beijing 100049, China}
\email{}

\author{Dingao Hu}
\affiliation{National Astronomical Observatories, Chinese Academy of Sciences, 20A Datun Road, Beijing 100101, China}
\affiliation{School of Astronomy and Space Sciences, University of Chinese Academy of Sciences(UCAS), \\Yuquan Road NO.19A Beijing 100049, China}
\email{huda@bao.ac.cn}

\author{Hengjie Lin}
\affiliation{National Astronomical Observatories, Chinese Academy of Sciences, 20A Datun Road, Beijing 100101, China}
\email{}

\author{Furen Deng}
\affiliation{National Astronomical Observatories, Chinese Academy of Sciences, 20A Datun Road, Beijing 100101, China}
\affiliation{School of Astronomy and Space Sciences, University of Chinese Academy of Sciences(UCAS), \\Yuquan Road NO.19A Beijing 100049, China}
\email{}

\author{Xuelei Chen}
\affiliation{National Astronomical Observatories, Chinese Academy of Sciences, 20A Datun Road, Beijing 100101, China}
\affiliation{School of Astronomy and Space Sciences, University of Chinese Academy of Sciences(UCAS), \\Yuquan Road NO.19A Beijing 100049, China}
\affiliation{Department of Physics, College of Sciences, Northeastern University, Shenyang 110819, China}
\affiliation{Centre for High Energy Physics, Peking University, Beijing 100871, China}
\affiliation{State Key Laboratory of Radio Astronomy and Technology, China}
\email{xuelei@bao.ac.cn}

\begin{abstract}

We present a comprehensive forecast for cosmological constraints using the joint observation of the cosmic shear signal from the Chinese Space Station Survey Telescope (CSST) and the clustering signal from the next-generation gravitational wave (GW) detector networks, e.g. Einstein Telescope (ET) and Cosmic Explorer (CE). By leveraging the angular clustering of astrophysical gravitational wave sources (AGWS) from the third-generation detectors and CSST's weak lensing surveys, we develop a theoretical framework to compute auto- and cross-angular power spectra of AGWS clustering, cosmic shear, and their cross-correlation. Mock datasets are generated by considering the detector-specific selection functions, uncertainties in luminosity distance, and weak lensing systematics. We employ the Markov Chain Monte Carlo (MCMC) methods to constrain the $\Lambda \mathrm{CDM}$ cosmological parameters, AWGS bias parameters, and star formation rate (SFR) parameters under three detector configurations. Our results demonstrate that the joint observation can achieve sub-$5\%$ precision on $H_0$ ($2.19\%$) and $w$ ($5.7\%$). Besides, the AGWS clustering bias parameters can be constrained to the precision of $4\%-5\%$, enabling the differentiation between stellar-origin compact binaries and primordial black hole scenarios. This multi-messenger approach also can be helpful to resolve mass-redshift degeneracies in the dark siren methods, providing independent validation for Hubble tension. Our work indicates that the joint observation of the third-generation GW detectors and the CSST can be a powerful probe of the large-scale structure and the cosmic expansion history.
\end{abstract}


\section{Introduction} 
Gravitational waves (GW), first predicted in Einstein's general relativity \citep{1-1916SPAW.......688E,2-1918SPAW.......154E}, arise from accelerated massive objects analogous to electromagnetic radiation from accelerated charges. Indirect evidence was given by the orbital period decay observations of binary neutron star systems \citep{3-1982ApJ...253..908T}, where energy losses via gravitational wave emission as predicted by the theory. The direct detection breakthrough occurred in 2015 with the GW150914 event detected by Advanced LIGO \citep{4-PhysRevLett.116.061102,5-2015,6-2010CQGra..27h4006H}, marking the beginning of gravitational wave astronomy. Subsequent joint observations of Advanced LIGO-Virgo networks have cataloged both binary black hole mergers and the GW170817 neutron star collision \citep{7-abbott2017multi}, which synergized electromagnetic and neutrino counterparts to pioneer multi-messenger astrophysics \citep{8-arnett1989supernova}.

Gravitational wave research and detection can be broadly categorized into two classes based on source distinguishability: resolvable transients (e.g. compact binary coalescences and core-collapse supernovae) and stochastic gravitational wave background (SGWB) \citep{9-christensen2018stochastic}. The SGWB is further classified into the astrophysical gravitational wave background (AGWB), comprising cumulative signals from various astrophysical sources including compact binary coalescences \citep{10-farmer2003gravitational}, primordial black holes \citep{11-vanzan2024gravitational}, asymmetric pulsars \citep{12-regimbau2007stochastic}, and supermassive black holes, and the cosmological stochastic gravitational wave background (CGWB), originating from violent processes in the early universe such as primordial density perturbations \citep{13-ananda2007cosmological}, cosmic inflation \citep{14-easther2008gravitational}, and first-order phase transitions \citep{15-caprini2009stochastic}.

The astrophysical gravitational wave sources (AGWS) makes an important relation between galactic cosmology and compact object studies \citep{24-cusin2018first}, since the AGWS can trace matter distribution through their clustering properties like galaxies \citep{31-oguri2016measuring}. Hence, the detection of AGWS clustering will impose stringent constraints on compact object physics (e.g. equation of state and spin distributions), initial mass function, and cosmic star formation history \citep{25-regimbau2011astrophysical}, enable large-scale structure (LSS) investigations through anisotropy measurements, and provide novel constraints on cosmological parameters via angular power spectrum analysis \citep{23-namikawa2016anisotropies,26-contaldi2017anisotropies,27-jenkins2018anisotropies,28-zheng2023angular}.

The cross-correlation between AGWS with LSS tracers such as galaxies represents a transformative multi-messenger probe \citep{118-2023JCAP...06..050B,119-Mukherjee:2022afz}, which can effectively extract cosmological information \citep{116-PhysRevD.103.043520,120-Sala:2025wwu}. In addition, this approach enables the measurement of the AGWS clustering bias \citep{112-2025MNRAS.537.1912Z}, with the tomographic analysis of the cross-correlation signal across redshift bins, which could  reveal the evolution of the bias and offer a direct test of structure formation and merger rate evolution \citep{113-2020PhRvR...2b3314C}. The inclusion of relativistic effects such as magnification further allows tests of gravity on cosmological scales and provides independent constraints on the AGWS population properties \citep{114-2020MNRAS.494.1956M,115-article,117-2024MNRAS.534.1283A}.

On the other hand, the weak gravitational lensing survey provides a direct probe of the projected matter distribution and its growth over cosmic time. In modern cosmological analyses, this signal is quantified via the cosmic shear angular power spectrum (or its real-space counterpart, the two-point correlation function), which is sensitive primarily to the parameters of amplitude of matter fluctuations $\sigma_8$ and the matter density $\Omega_m$ \citep[e.g.][]{29-bartelmann2001weak}. The Stage-III and Stage-IV surveys always employ a multi-probe $3\times2\rm{pt}$ framework \citep{107-2021arXiv210513548K,108-2023A&A...675A.189D,109-2024} that jointly analyzes three two-point functions of weak lensing and galaxy clustering, which can extract cosmological information accurately and break the degeneracies between parameters, such as DES Y6 \citep{110-2026arXiv260114559D} and DESI DR1 $3\times2\rm{pt}$ analyses \citep{111-2025arXiv251215960P}.

Compared to the AGWS--galaxy cross-correlation, the cross-correlation between AGWS and weak lensing can directly link the spatial distribution of GW events to the underlying matter field. Such a measurement is sensitive not only to the expansion history but also to the growth of structure. Consequently, the cross-correlation between AGWS and weak lensing can deliver novel and self-consistent cosmological constraints, which would provide complementary insights to address current cosmological issues, such as the Hubble tension\citep{32-riess2021cosmic,33-2014A&A...571A..16P,34-hazra2015post,35-novosyadlyj2014constraining,36-luis2016trouble}, and explore the properties of dark matter, dark energy, and AGWS population.

For observation, the future third-generation ground-based gravitational wave detectors, e.g. Cosmic Explorer (CE) \citep{43-Abbott_2017} and Einstein Telescope (ET) \citep{44-punturo2010einstein,45-hild2011sensitivity}, exhibit significantly enhanced sensitivity and detection capabilities in the $10^0-10^3\; \mathrm{Hz}$ frequency band, compared to the current instruments like Advanced LIGO and Advanced Virgo. Joint observations with these third-generation detectors are expected to detect $10^4-10^6$ binary neutron star (BNS) and binary black hole (BBH) mergers, with sky localization accuracy reaching $1\;\mathrm{deg}^2$ at $90\%$ confidence level for approximately $10^3$ BNS events and $10^4$ BBH events \citep{46-borhanian2024listening}. Meanwhile, the Chinese Space Station Survey Telescope (CSST) scheduled for launch in 2027, will conduct $17500 \; \mathrm{deg^2}$ photometric imaging and slitless spectroscopic surveys \citep{91-ZhanHu2011, 92-ZhanHu2021, 47-gong2019cosmology, 93-csstcollaboration2025introductionchinesespacestation, 94-2025SCPMA..6880402G}, which can precisely measure the signals of weak gravitational lensing and galaxy clustering for the LSS analysis. 

In this work, we investigate the cross-correlation of the CSST shear measurement and the AGWS clustering from the third-generation ground-based gravitational wave detectors (i.e. ET and CE) and sensitivity-upgraded Advanced LIGO and Virgo \citep{5-2015, 96-Abbott_2020} and KAGRA \citep{97-2019}. By theoretically generating the mock data of the AGWS clustering, AGWS-shear, and shear power spectra, we employ the Markov Chain Monte Carlo (MCMC) method to explore the constraining power of joint observations on the cosmological parameters, cosmic star formation rate (SFR), and other parameters related to gravitational wave sources.

The paper is organized as follows: in Section \ref{Methodology} we provide a detailed exposition of our theoretical framework of the AGWS clustering and weak lensing. In Section \ref{Observation}, we present the generation of the mock data from the three gravitational wave detector networks along with the CSST. The model fitting and constraint results are given in Section \ref{Result}, and our conclusions are summarized in Section \ref{Conclusion}.

\section{Methodology}\label{Methodology}
\subsection{AGWS clustering}

\begin{figure*}[htbp] 
    \centering
    \begin{minipage}{0.48\textwidth}
        \centering
        \includegraphics[width=\linewidth]{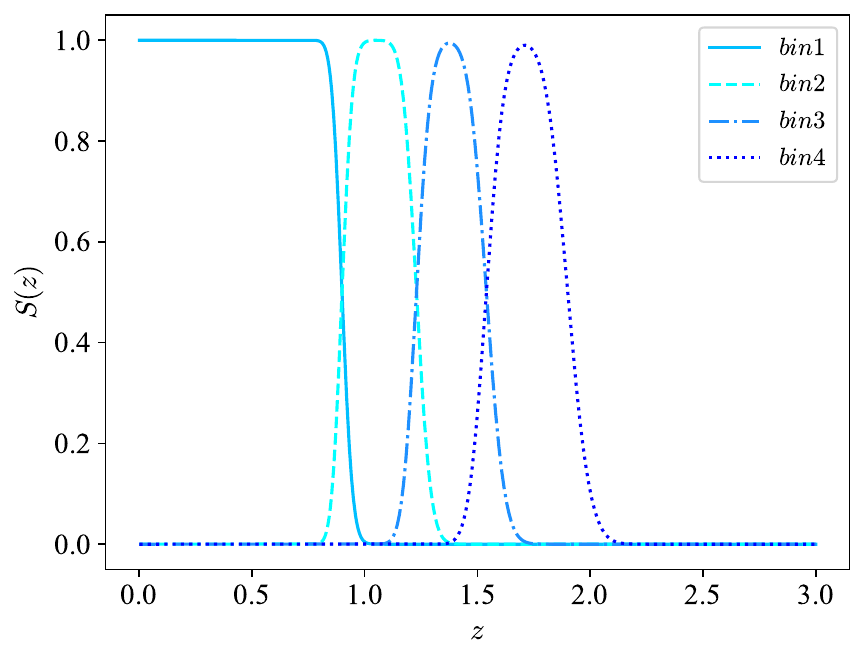}
    \end{minipage}
    \hfill 
    \begin{minipage}{0.48\textwidth} 
        \centering
        \includegraphics[width=\linewidth]{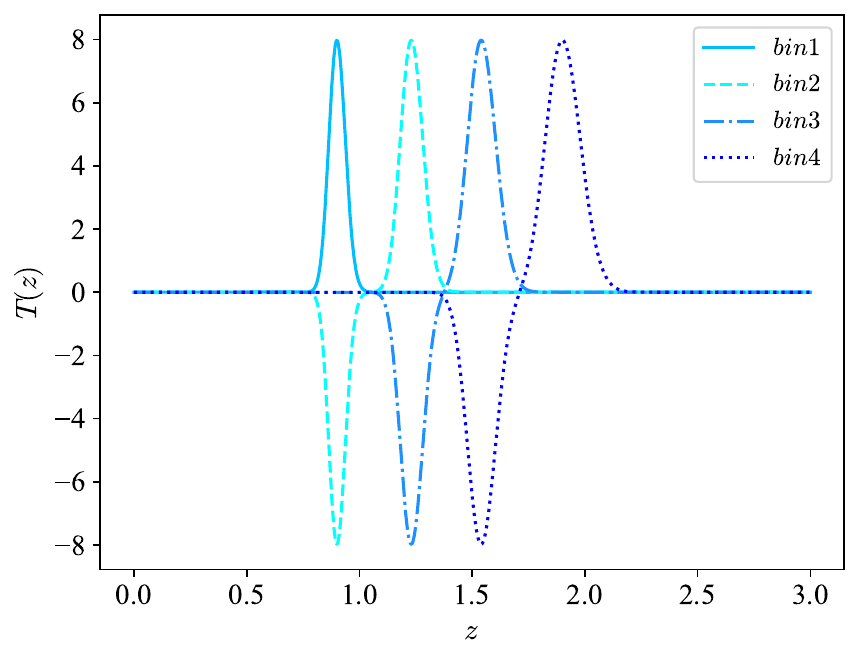}
    \end{minipage}
    \caption{The functions of $S_i(z)$ (left panel) and $T_i(z)$ (right panel), based on the ECS configuration of the GW detector network.}
    \label{SF}
\end{figure*}

As we know, the luminosity distance of compact binary gravitational wave sources can be inferred from their GW waveforms \citep{98-1986Natur.323..310S}. However, the observed luminosity distance often has deviations from its true value due to multiple factors, including statistical uncertainties in GW observations, parameter degeneracies between luminosity distance, and other parameters such as the component masses and inclination angle of the compact binaries. Here we assume that the relationship between the observed luminosity distance $D_{\rm L,obs}$ from the analysis of the waveform and the true value $D_{\rm L}$ can be described by a log-normal distribution, which is given by \citep{31-oguri2016measuring}
\begin{align}
    p(D_{\rm L, obs} | D_{\rm L}) =&\ \frac{1}{\sqrt{2\pi}\sigma_{{\rm ln}D}D_{\rm L,obs}} \nonumber \\
    &\times \exp\left[-x^2(D_{\rm L,obs},D_{\rm L})\right], 
\end{align}
where
\begin{equation}
    x(D_{\rm L,obs},D_{\rm L}) = \frac{\ln(D_{\rm L,obs}) - \ln(D_{\rm L})}{\sqrt{2}\sigma_{{\rm ln}D}}.
\end{equation}

Here we adopt $\sigma_{{\rm ln}D}=0.05$ \citep{48-PhysRevLett.110.151103}. During their propagation from a source at redshift $z$ to Earth, gravitational waves are influenced by the LSS of the Universe, specifically through the weak gravitational lensing effect \citep{49-Takahashi_2005}. Due to the weak gravitational lensing effect, the observed luminosity distance to a source is statistically modified relative to the value predicted in a homogeneous and isotropic Friedmann-Lemaitre-Robertson-Walker (FLRW) cosmological model. This modification arises from the integrated matter distribution along the line of sight, which can be expressed as
\begin{equation}
    {D_{\rm L}} = \bar{D}_{\rm L} \times \mu^{-\frac{1}{2}}(\bm{\theta},z) \approx \bar{D}_{\rm L} \times [1-\kappa(\bm{\theta},z)],
\end{equation}
where $\bar{D}_{L}$ denotes the luminosity distance calculated in a homogeneous and isotropic FLRW universe, where $\mu$ is the magnification. Under the weak lensing approximation, we have $\mu \approx 1+2\kappa$, and $\kappa(\bm{\theta},z)$ is the convergence field that characterizes the projection of the matter density field $\delta_m(\bm{\theta},z)$ along the line of sight in the direction $\bm{\theta}$, which is expressed as \citep{29-bartelmann2001weak}
\begin{align}
    \kappa(\bm{\theta},z) =& \int_o^z dz' W^\kappa(z';z)\delta_{\rm m}(\bm{\theta},z')= \frac{3\Omega_{\rm m}H_0^2}{2c} \nonumber \\ 
    & \times \int_0^z dz' \frac{1+z'}{H(z')} \frac{D_{\rm A}(z',z)D_{\rm A}(z')}{D_{\rm A}(z)}\delta_{\rm m}(\bm{\theta},z').
\end{align}
Here $W^\kappa(z';z)$ denotes the gravitational lensing weight function, $\delta_{\rm m}(\bm{\theta},z)$ represents the matter density contrast, $\Omega_{\rm m}$ is the dimensionless matter density parameter, $H_0=100h$ is the Hubble constant, $c$ is the speed of light, and $H(z)$ represents the Hubble parameter. $D_{\rm A}(z)$ denotes the comoving angular diameter distance, which is defined as
\begin{equation}
    D_{\rm A}(z)=c\int_a^1\frac{da}{a^2H(a)}=c\int_0^z\frac{dz}{H(z)},
\end{equation}
where $a=1/(1+z)$ is the scale factor.

All gravitational wave sources can be categorized into different bins based on the luminosity distance, and their redshifts can be derived by assuming or giving a specific cosmology. The angular number density of gravitational wave sources $n_i^{gw}(\bm{\theta})$ in the $i$-th bin can be derived by projecting the 3D number density $n^{gw}(\bm{\theta},z)$ as \citep{31-oguri2016measuring}
\begin{equation}
n_i^{gw}(\bm{\theta}) = \int_0^\infty dz \frac{cD_{\rm A}^2(z)}{H(z)}G_i(\bm{\theta},z)\,n^{gw}(\bm{\theta},z), \label{eq:n_gw}
\end{equation}
where $G_i(\bm{\theta},z)$ denotes the selection function of the $i$-th bin along the line of sight. Given the weak influence of gravitational lensing on luminosity distance measurements, we expand $G_i(\bm{\theta},z)$ around $\bar{D}_L$ via Taylor series:
\begin{align}
G_i(\bm{\theta},z) 
=&\ \frac{1}{2}\Bigl[\erfc\bigl(x(D_{\rm L}(\bm{\theta},z),D_{i,\min})\bigr) \nonumber \\ 
& - \erfc\left(x\bigl(D_{\rm L}(\bm{\theta},z),D_{i,\max}\right)\bigr)\Bigr] \nonumber \\
\approx&\ G_i\left(\bar{D}_{\rm L}(z)\right) + \left.\frac{\partial G_i}{\partial D_{\rm L}}\right|_{D_{\rm L}=\bar{D}_L}(D_{\rm L}-\bar{D}_{\rm L}) \nonumber \\
=&\ S_i(z) + \kappa(\bm{\theta},z)T_i(z),
\end{align}
where $S_i(z)$ and $T_i(z)$ are defined as
\begin{align}
    S_i(z) =&\ \frac{1}{2}\Bigl[\erfc\bigl(x(\bar{D}_{\rm L}(z),D_{i,\min})\bigr) \nonumber \\
    & -\erfc\left(x\bigl(\bar{D}_{\rm L}(z),D_{i,\max}\right)\bigr)\Bigr], \\
    T_i(z) =&\ \frac{1}{\sqrt{2\pi}\sigma_{{\rm ln}D}}\Bigl[-\exp\left(-x(\bar{D}_{\rm L}(z),D_{i,\min})\right) \nonumber \\
    & +\exp\left(-x(\bar{D}_{\rm L}(z),D_{i,\max})\right)\Bigr].
\end{align}
The corresponding $S_i(z)$ and $T_i(z)$ are shown in Figure~\ref{SF}, when assuming the parameters of the ECS configuration (the details can be found in Section~\ref{GW Observation}).

\begin{figure*}[htbp] 
    \centering
    \begin{minipage}{0.48\textwidth}
        \centering
        \includegraphics[width=\linewidth]{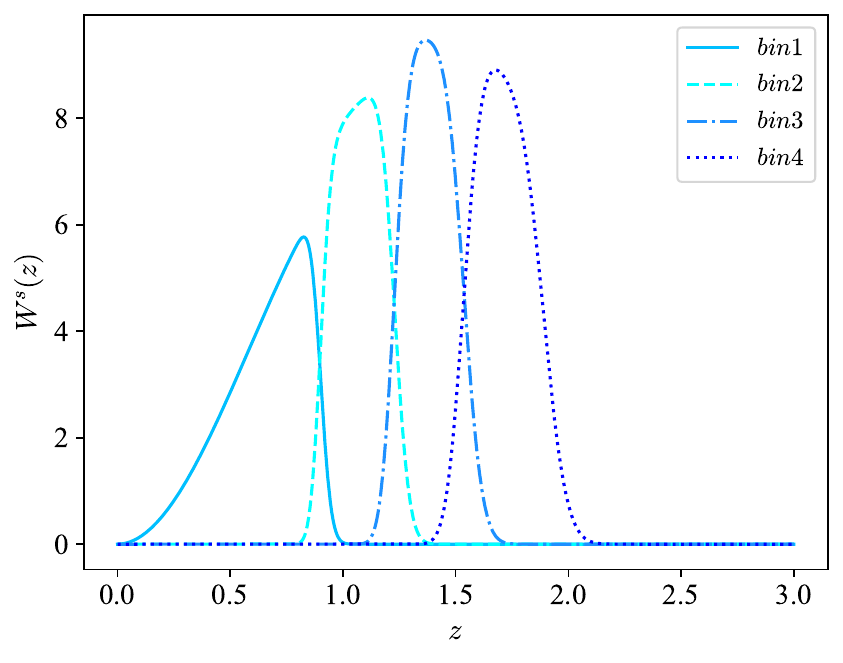} 
    \end{minipage}
    \hfill 
    \begin{minipage}{0.50\textwidth} 
        \centering
        \includegraphics[width=\linewidth]{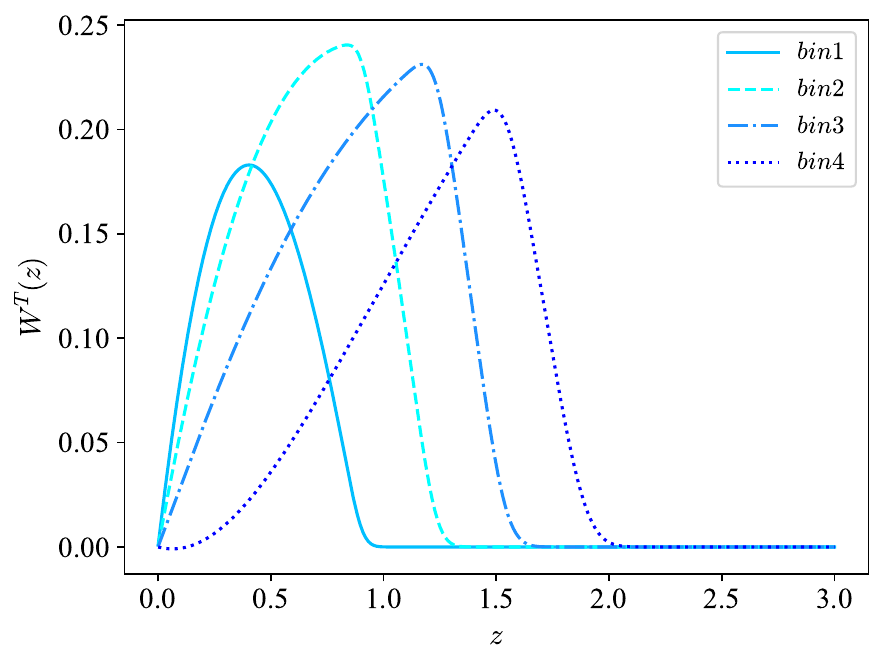}
    \end{minipage}
    \caption{The weighting functions of $W_i^S(z)$ (left panel) and $W_i^T(z)$ (right panel), based on the ECS configuration of the GW detector network.}
    \label{WF_GW}
\end{figure*}

By averaging the solid angle $\bm{\theta}$ in Equation~(\ref{eq:n_gw}) over the full sky, the mean projected number density for the $i$-th bin is
\begin{align}
\bar{n}_i^{gw} \approx& \int_0^\infty dz \frac{cD_{\rm A}^2(z)}{H(z)}{S_i(z)}\,{\bar{n}^{gw}(z)} \nonumber \\
=& \int_0^\infty dz \frac{cD_{\rm A}^2(z)}{H(z)}S_i(z)\, T_{\rm obs}\frac{\dot{n}^{gw}(z)}{1+z}, \label{nbar_gw}
\end{align}
where $T_{\rm obs}$ denotes the observational time of GW detectors, and $\dot{n}^{gw}(z)$ represents the detectable merger rate, calculated as the product of detector efficiency $E_{\rm det}(z)$ (the details can be found in Section~\ref{GW Observation}) and intrinsic merger rate $R^{gw}(z)$, i.e. $\dot{n}^{gw}(z) = E_{\rm det}(z)\times R^{gw}(z)$. The $1+z$ factor accounts for cosmological time dilation. The first approximate equality in Equation~(\ref{nbar_gw}) is due to the neglect of the second term in $G_i(z)$, as it is much smaller compared to the first term $S_i(z)$.

For modeling the intrinsic merger rate $R^{gw}(z_{\rm m})$ at merger redshift $z_{\rm m}$, we adopt the convolution between cosmic SFR and delay time distribution $P(t_{\rm d})$ \citep{50-DE_FREITAS_PACHECO_2006,51-Regimbau_2009,52-Banerjee_2009,53-O_Shaughnessy_2010}, by assuming the number density of gravitational-wave sources at redshift $z$ is proportional to the SFR$(z)$, which is reasonable for compact binaries. Then we have
\begin{align}
    {R^{gw}(z_{\rm m})} &= A \int_{t_{\rm d,min}}^{t_{\rm d,max}} dt_{\rm d}\, {\rm SFR}({z_{\rm f}})\,P(t_{\rm d}) \nonumber \\
    &= A \int_{z_{\rm f,min}}^{z_{\rm f,max}} dz_{\rm f}\, \frac{dt_{\rm f}}{dz_{\rm f}}\, {\rm SFR}({z_{\rm f}})\, P(t_{\rm f}-t_{\rm m}).
    \label{eq:R_gw}
\end{align}
Here  $t_{\rm m}$, $t_{\rm f}$, and $t_{\rm d}$ are the merger time, binary star formation time, and delay time, respectively. The delay time is defined as the time interval between the formation of a binary stellar system and its eventual merger as a compact binary, which follows a specific distribution and satisfy $t_{\rm d} = t_{\rm f} - t_{\rm m}$, all timescales refer to cosmological lookback time $t= \int_0^z \frac{dz'}{(1+z')H(z')}$. The delay time distribution $P(t_{\rm d})$ follows a power-law form from population synthesis models \citep{53-lipunov1995evolution,50-DE_FREITAS_PACHECO_2006,55-belczynski2006study}, which is given by
\begin{equation}
    P(t_{\rm d}) \propto t_{\rm d}^{-1} = \frac{1}{\ln(t_{\rm d,max}/t_{\rm d,min})t_{\rm d}}.
\end{equation}
The minimum delay times are set as $t_{\rm d,min}=20\,\mathrm{Myr}$ \citep{51-Regimbau_2009,46-borhanian2024listening} for BNS and BHNS systems, and $10\, \mathrm{Myr}$ \citep{56-Vitale_2019} for BBH systems. The maximum delay time is  $t_{\rm d,max}={\rm min}\left\{10\,\mathrm{Gyr},t(z; 20)\right\}$, where $t(z; 20)$ denotes the time interval between redshift $z$ and $z=20$. The normalization constant $A$ ensures $R^{gw}(z=0)$ matches observed local merger rates from the GWTC-3 results \citep{58-abbott2023population}.

The SFR we adopt follows the Madau-Dickinson model \citep{59-madau2014cosmic}, which can be expressed as
\begin{equation}
{\rm SFR}(z) = \psi_0 \frac{(1+z)^{\alpha}}{1+\left[(1+z)/C\right]^{\beta}},
\end{equation}
where $\alpha=2.7$, $\beta=5.6$, and $C=2.9$ \citep{56-Vitale_2019}. Note that  the $\psi_0=0.015\;\mathrm{M_{\odot}year^{-1} Mpc^{-3}}$ can be absorbed by the normalization constant $A$ in Equation~(\ref{eq:R_gw}).

Similar to galaxy clustering, we characterize the clustering of AGWS using the relative fluctuation in their projected number density. The primary perturbations originate from inhomogeneities in the 3D spatial distribution of gravitational wave sources $n^{gw}(\bm{\theta},z)=[1+\delta_{gw}(\bm{\theta},z)]\,\bar{n}^{gw}(z)$, where $\bar{n}^{gw}(z)$ is the average spatial number density at redshift $z$, and the secondary contribution is from weak gravitational lensing. We define the relative density fluctuation in the $i$-th bin as
\begin{align}
    \delta_i^{gw}(\bm{\theta}) &= \frac{n_i^{gw}(\bm{\theta}) - \bar{n}_i^{gw}}{\bar{n}_i^{gw}} \nonumber \\
    &\approx \frac{1}{\bar{n}_i^{gw}}\int_0^\infty dz \frac{cD_{\rm A}^2(z)}{H(z)}S_i(z)\delta^{gw}(\bm{\theta},z) \nonumber \\  
    &\quad +\frac{1}{\bar{n}_i^{gw}}\int_0^\infty dz \frac{cD_{\rm A}^2(z)}{H(z)}T_i(z)\kappa(\bm{\theta},z) \bar{n}^{gw}(z).
\end{align}
We assume $\delta^{gw}(\bm{\theta},z)=b^{gw}(z)\delta_{\rm m}({\bm \theta},z)$, where $b^{gw}$ is the AGWS bias. We adopt a simple form $b^{gw}(z)=1+1/D(z)$ \citep{31-oguri2016measuring} , and D(z) is the linear growth factor. Note that, when generating the mock data, we treat the bias in each tomographic bin as a constant calculated using this formula (see Section \ref{GW Observation} for details), and set it as a free parameter in the fitting process. Then $ \delta_i^{gw}(\bm{\theta})$ can be expressed as
\begin{equation}
    \delta_i^{gw}(\bm{\theta}) = \int_0^{\infty}dz \bigl[W_i^{S}(z)+W_i^{T}(z) \bigr]\delta_{\rm m}(\bm{\theta},z),
\end{equation}
where $W_i^{S}(z)$ and $W_i^{T}(z)$ represent the weight functions for gravitational wave source clustering and weak lensing effects, respectively. They take the forms as
\begin{align}
    W_i^S(z) = \frac{1}{\bar{n}_i^{gw}}\frac{cD_{\rm A}^2(z)}{H(z)}S_i(z) \bar{n}^{gw}(z)b_{gw}(z), \label{W_S}
\end{align}
\begin{align}
    W_i^T(z) = \frac{1}{\bar{n}_i^{gw}}\int_{z}^{\infty}\frac{cD^2_A(z')}{H(z')}W^\kappa(z;z')T_i(z')\bar{n}_{gw}(z')dz'. \label{W_T}
\end{align}
 We show $W_i^{S}(z)$ and $W_i^{T}(z)$ in Figure~\ref{WF_GW}. As can be seen, $W_i^S(z)$ exhibits only partial overlap in adjacent redshift bins, whereas $W_i^T(z)$ shows overlapping profiles across all redshift bins, and $W_i^T(z)$ is an order of magnitude smaller overall compared to $W_i^S(z)$. 

\begin{figure*}[htbp] 
    \centering
    \begin{minipage}{0.48\textwidth}
        \centering
        \includegraphics[width=\linewidth]{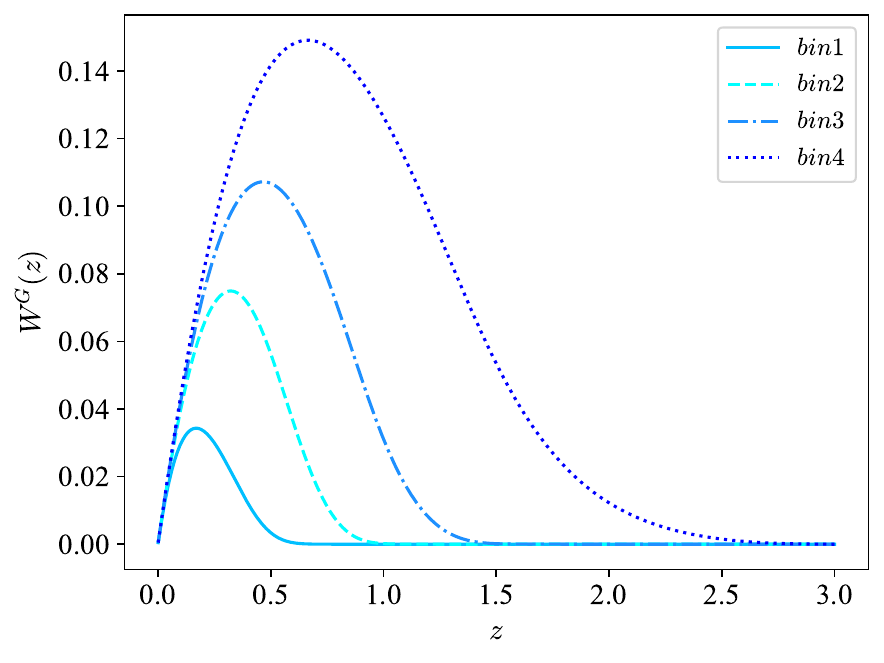}     \end{minipage}
    \hfill 
    \begin{minipage}{0.48\textwidth} 
        \centering
        \includegraphics[width=\linewidth]{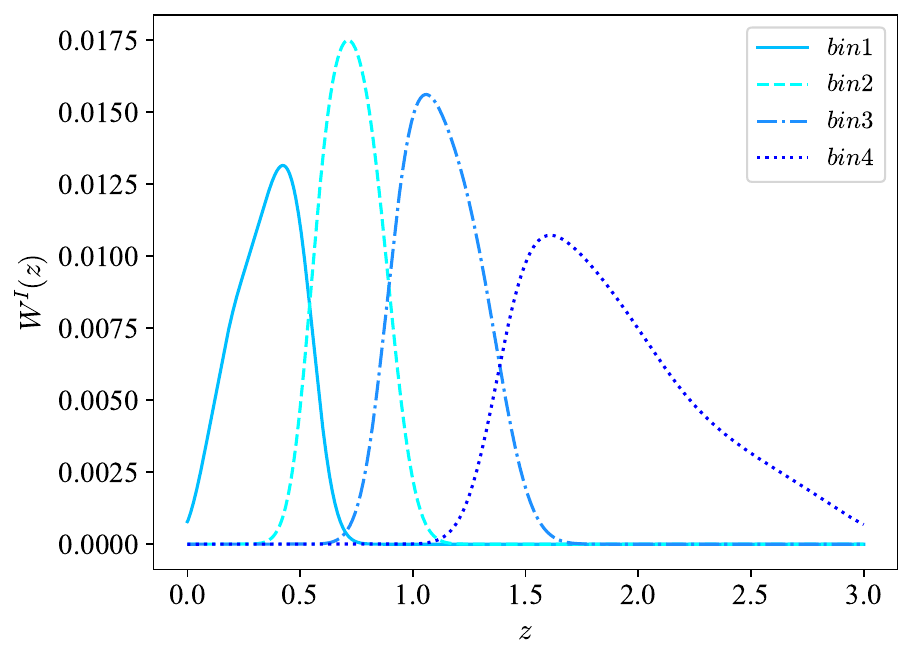}    
    \end{minipage}
    \caption{The weighting functions $W_i^G(z)$ (left panel) and $W_i^I(z)$ (right panel), based on the ECS configuration of the GW detector network.}
    \label{WF_WL}
\end{figure*}

\subsection{Weak lensing}
Weak gravitational lensing can be characterized by the convergence effect \citep{29-bartelmann2001weak,61-Kilbinger_2015}. While convergence $\kappa(\bm{\theta},z)$ inherently represents a projection of matter density field, actual sky surveys involve broad source redshift distributions, making observables weighted averages over source distributions in a distance or redshift $i$-th tomographic bin, i.e. the so-called lensing tomography technique \citep{62-Hu_1999}. It gives  
\begin{align}
\kappa_i^{G}(\bm{\theta}) &= \int_0^{\infty}dz\, n_i^g(z)\,\kappa(\bm{\theta},z) \nonumber \\
&= \int_0^{\infty}dz\, W_i^{G}(z)\,\delta_{\rm m}(\bm{\theta},z),
\end{align}
where the weight $W_i^G(z)$ is defined as
\begin{equation}
    W_i^{G}(z) = \frac{3\Omega_{\rm m}H_0^2}{2c}\int_z^{\infty} dz' \frac{1+z}{H(z)} \frac{D_{A}(z,z')D_{A}(z)}{D_{A}(z')} n_i^g(z).\label{W_G}
\end{equation}
Here $n_i^g(z)$ denotes the normalized galaxy distribution function for the $i$-th bin, which satisfies $\int_0^{\infty}dz\, n_i^g(z) = 1$.
Besides, we also consider the Intrinsic Alignment (IA) effect \citep{63-Troxel_2015}, which can be estimated by
\begin{align}
    \kappa_i^I(\bm{\theta})=\int_0^{\infty}dz W_i^{I}(z) \delta_{\rm m}(\bm{\theta},z), 
    \end{align}
where the weight $W_i^I(z)$ is defined as
\begin{align}
    W_i^I(z)=A_{\rm IA}C_1\rho_{c0}\frac{\Omega_{\rm m}}{D(z)}\left(\frac{1+z}{1+z_0}\right)^{\eta_{\rm IA}}\left(\frac{L_i}{L_0}\right)^{\beta_{\rm IA}}n_i^g(z). \label{W_I}
\end{align}
Here $C_1=5\times10^{-14}\;h^{-2}\mathrm{M_{\odot}^{-1}Mpc^{3}}$ is a normalization constant, $\rho_{c0}$ is the present critical density, and $z_0$ and $L_0$ are the pivot redshift and luminosity, respectively. $A_{\rm IA}$, $\eta_{\rm IA}$ and $\beta_{\rm IA}$ are free parameters. We neglect luminosity dependence \citep{64-2017MNRAS.465.1454H,65-2018MNRAS.474.4894J}, and fix $\beta_{\rm IA}=0$. We adopt $z_0=0.6$, and set the fiducial values $A_{\rm IA}=1$, $\eta_{\rm IA}=0$ in this work. Then the total convergence becomes
\begin{align}
    \kappa_i(\bm{\theta}) &= \kappa_i^G(\bm{\theta}) + \kappa_i^I(\bm{\theta}).
\end{align}

The weighting functions $W_i^{G}(z)$ and $W_i^{I}(z)$ are shown in Figure~\ref{WF_WL}. It illustrates that $W_i^{G}(z)$ at higher redshift bins encompasses the distribution at lower redshift bins, while $W_i^{I}(z)$ shows only partial overlap between adjacent redshift bins, and $W_i^{G}(z)$ is dominant in the total convergence as expected.

\subsection{Angular Power Spectrum}
The auto and cross angular power spectra for AGWS clustering and weak lensing are expressed as
\begin{align}
C_{ij}^{gwgw}(\ell) &= C_{ij}^{SS}(\ell)+C_{ij}^{ST}(\ell)+C_{ij}^{TS}(\ell)+C_{ij}^{TT}(\ell), \\
C_{ij}^{gwwl}(\ell) &= C_{ij}^{SG}(\ell)+C_{ij}^{TG}(\ell)+C_{ij}^{SI}(\ell)+C_{ij}^{TI}(\ell), \\
C_{ij}^{wlwl}(\ell) &= C_{ij}^{GG}(\ell)+C_{ij}^{GI}(\ell)+C_{ij}^{IG}(\ell)+C_{ij}^{II}(\ell).
\end{align}
The superscripts $S$, $T$, $G$ and $I$ denote that the angular power spectrum is computed using the corresponding weight functions $W^S(z)$, $W^T(z)$, $W^G(z)$ and $W^I(z)$ defined in Equations~(\ref{W_S}), (\ref{W_T}), (\ref{W_G}), (\ref{W_I}), respectively.
Using Limber approximation \citep{66-1954ApJ...119..655L}, we can compute the angular power spectra by
\begin{align}
    {C_{ij}^{AB}(\ell)} =& \int_0^{\infty}dz \frac{H(z)}{c}\frac{W_i^A(z)W_j^B(z)}{D_{\rm A}^2(z)} \nonumber \\
    &\times P_{\rm m}\left(k=\frac{\ell+1/2}{D_{\rm A}(z)},z\right),
\end{align}
where $A,B=\{S,T,G,I\}$, and $P_{\rm m}(k,z)$ is the matter power spectrum calculated by {\tt CAMB} \citep{99-2000ApJ...538..473L}. Given gravitational wave localization uncertainties, we adopt $\ell_{\rm max}=300$ \citep{100-2018PhRvD..98h3524O} (only for ECS, the corresponding $k_{\rm max}$ is between 0.09 and 0.18 ${\rm Mpc^{-1}}h$ at the effective redshifts of the tomographic bins we consider, see Section \ref{GW Observation} for details), and employ the linear matter power spectrum, neglecting the effects of neutrino and baryonic feedback.

Considering the shot noise and systematics, the total AGWS clustering power spectrum is
\begin{equation}
\hat{C}_{ij}^{gwgw}=C_{ij}^{gwgw}+\frac{\delta_{ij}}{\bar{n}_i^{gw}}+N_{\rm sys}^{gw},
\end{equation}
where $\delta_{ij}$ is the Kronecker delta, and $N_{\rm sys}^{gw}$ is the systematical noise.
The cross power spectrum can be written as
\begin{equation}
\hat{C}_{ij}^{gwwl}=(1+m_j){C}_{ij}^{gwwl}+N_{\rm sys}^{gwwl},
\end{equation}
where $m_i$ is the parameter accounting for the multiplicative error in weak lensing measurements \citep{67-troxel2018dark}, and $N_{\rm sys}^{gwwl}$ is the systematical noise for the cross power spectrum.
Finally, the weak lensing power spectrum is given by
\begin{equation}
\hat{C}_{ij}^{wlwl}=(1+m_i)(1+m_j){C}_{ij}^{wlwl}+\delta_{ij}\frac{\sigma_\gamma^2}{\bar{n}_i^g}+N_{ij}^{\rm add}(\ell).
\end{equation}
Here $\sigma_\gamma^2$ denotes per-component shear variance from intrinsic ellipticity and measurement errors, $\bar{n}_i^g$ is the mean galaxy angular density, and $N_{ij}^{\rm add}(\ell)$ is the additive error \citep{68-2006MNRAS.366..101H}. For simplicity, we model $\bar{N}_{ij}^{\rm add}$ as scale-independent constants across all tomographic bins \citep[e.g.][]{69-zhan2006cosmic}.

\section{Mock data} \label{Observation}
\subsection{GW observation} \label{GW Observation}
Current gravitational wave detectors like LIGO, Virgo and KAGRA have provided substantial insights into compact binary populations \citep{70-Fishbach_2021,71-El_Bouhaddouti_2025,58-abbott2023population}, potential primordial origins of observed black holes \citep{72-Sasaki_2016,73-Clesse_2017}, and the stochastic gravitational wave background from BBH/BNS mergers \citep{74-2018PhRvL.120i1101A,75-2024arXiv241218318B}. In the next five years, Advanced LIGO/Advanced Virgo, along with new detectors KAGRA \citep{76-2013PhRvD..88d3007A} and LIGO-India \citep{77-2013IJMPD..2241010U}, will significantly enhance our understanding of physics and astronomy. 

\begin{figure}[t]
    \centering
    \includegraphics[width=1\linewidth]{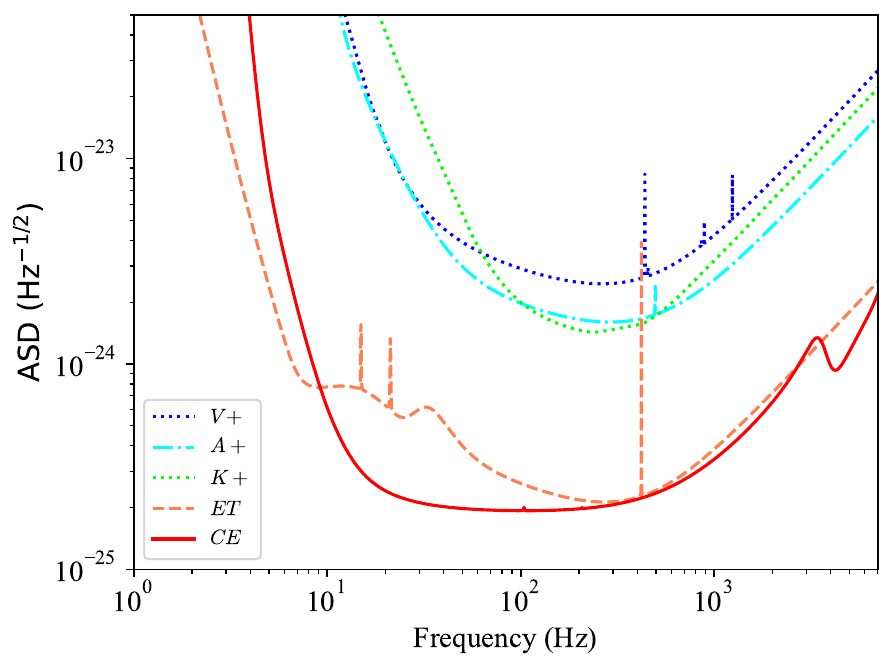}
    \caption{\label{fig_sensity}The ASD curves of the two generation GW detectors, including Advanced-plus upgrade of LIGO (A+: Hanford, Livingston, and India), Virgo (V+: Italy), and KAGRA (K+: Japan), and the third generation detectors Einstein Telescope (ET) and Cosmic Explorer (CE). The ET ASD is scaled by a factor of $2/3$ to represent the effective sensity of ET's triangular design, which is used in our analysis.}
\end{figure}

\begin{table*}[htbp]
    \centering
    \caption{The parameter values for the three GW detector networks. The values originates from numerical simulations in the referenced study \citep{46-borhanian2024listening}. The second column, $R_{\rm det}$, indicates the annual detection count of BNS and BBH events with $\rm SNR>10$. The third column $z_{50}$ denotes the redshift where detection efficiency drops to $50\%$, reflecting the network's detection depth. The last three columns $a,b,c$ specify parameters of the sigmoid function used for detector efficiency fitting.}
    \label{table_detec}
    \begin{tabular*}{\hsize}{@{}@{\extracolsep{\fill}}lccccc@{}}
    \toprule\toprule
              \multicolumn{6}{c}{\textbf{BNS}}                              \\
    \midrule
Network Name & $R_{\rm det}$    &  ${z_{50}}$      & $a$      & $b$        & $c$        \\
    \midrule
ECS          & 240000 & 2      & 5.834  & 0.00928  & 0.09996  \\
VKI+C        & 110000 & 1      & 5.471  & 0.1155   & 0.2049   \\
HLKI+E       & 41000  & 0.63   & 14.95  & 0.006631 & 0.1549   \\
    \midrule
              \multicolumn{6}{c}{\textbf{BBH}}                              \\
    \midrule
ECS          & 120000 & 34    & 0.1211  & 0.4108   & 0.2391   \\
VKI+C        & 110000 & 14    & 14.2    & 0.03823  & 0.003638 \\
HLKI+E       & 97000  & 7     & 14.2    & 5.08e-5  & 0.007811 \\
    \bottomrule
    \end{tabular*}
\end{table*}

However, the number of detectable compact binaries remains insufficient for clustering study, necessitating next-generation ground-based detectors. Third-generation facilities like Einstein Telescope and Cosmic Explorer, currently under development, plan to complete initial construction within $10-15$ years. In Figure~\ref{fig_sensity}, we show the amplitude spectral density (ASD) curves of the three GW detection\footnote{https://dcc.cosmicexplorer.org/public/0163/T2000007/005/}. These third-generation detectors will achieve dramatically improved sensitivity, enabling detection of $10^4-10^5$ merging compact binaries over $3-5$ years of observation \citep{46-borhanian2024listening} .

Since multi-detector networks can significantly improve the accuracy of source localization, here we consider three detector network configurations  \citep{46-borhanian2024listening}: {
\begin{itemize}
\item ECS: The standard configuration comprising ET, CE, and CE-South, which is expected to begin operations within 15-20 years, offering optimal sensitivity and localization precision.
\item VKI+C: A network combining three advanced-plus sensitivity detectors (Virgo, KAGRA, LIGO-India) with CE.
\item HLKI+E: A network consisting of ET, KAGRA and three Advanced-plus upgraded LIGO detectors (Hanford, Livingston, India).
\end{itemize}
}

\begin{figure}
    \centering
    \includegraphics[width=1\linewidth]{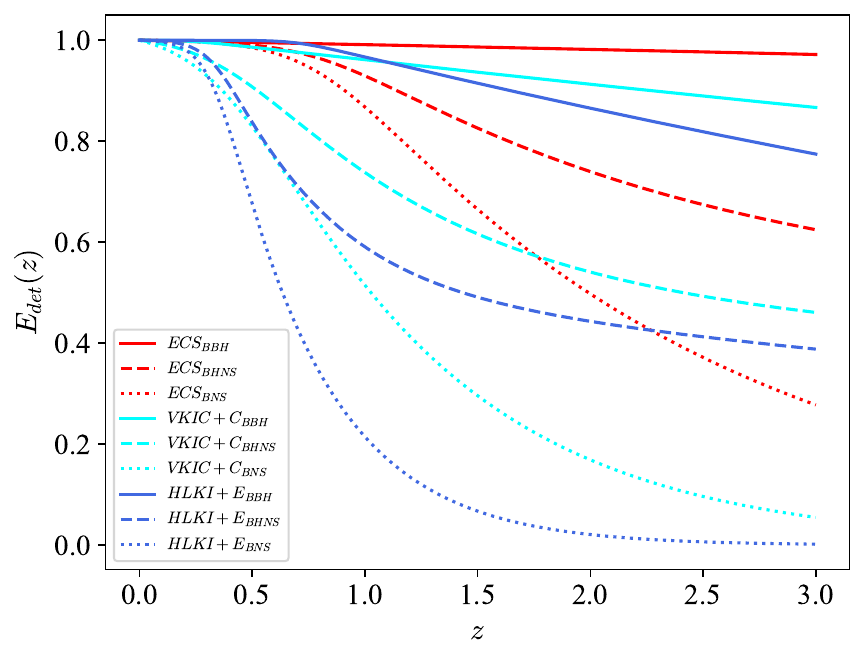}
    \caption{Detection rate comparison for the three detector networks. For BBH and BNS systems, we employ three-parameter sigmoid function fits (as listed in Table \ref{table_detec}) to model their detection rates.}
    \label{fig_DetectionRate}
\end{figure}

\begin{figure*}[t]
\epsscale{1.9}
\centerline{
\resizebox{!}{!}{\includegraphics[scale=0.4]{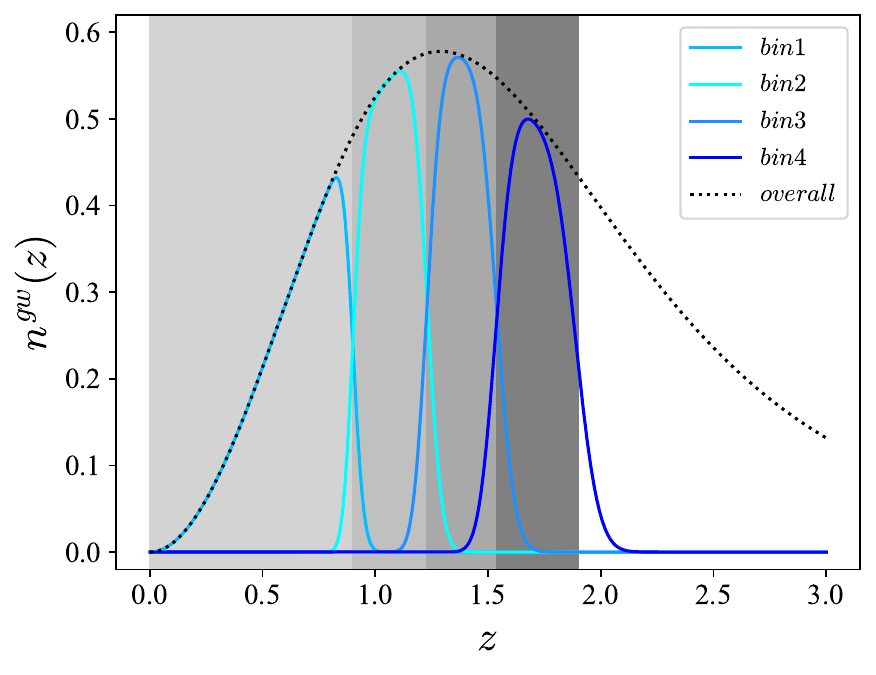}}
\resizebox{!}{!}{\includegraphics[scale=0.4]{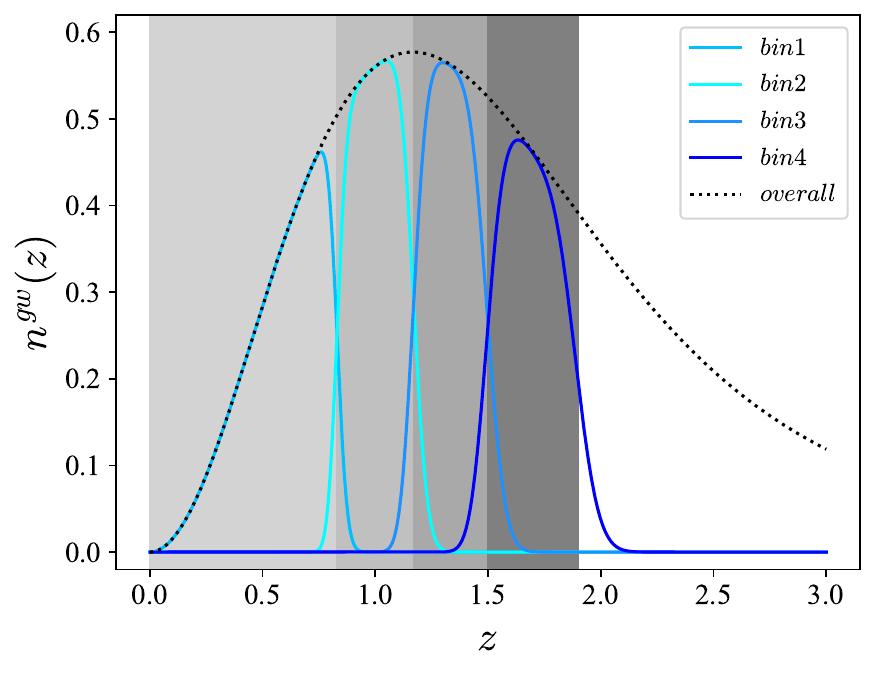}}
\resizebox{!}{!}{\includegraphics[scale=0.4]{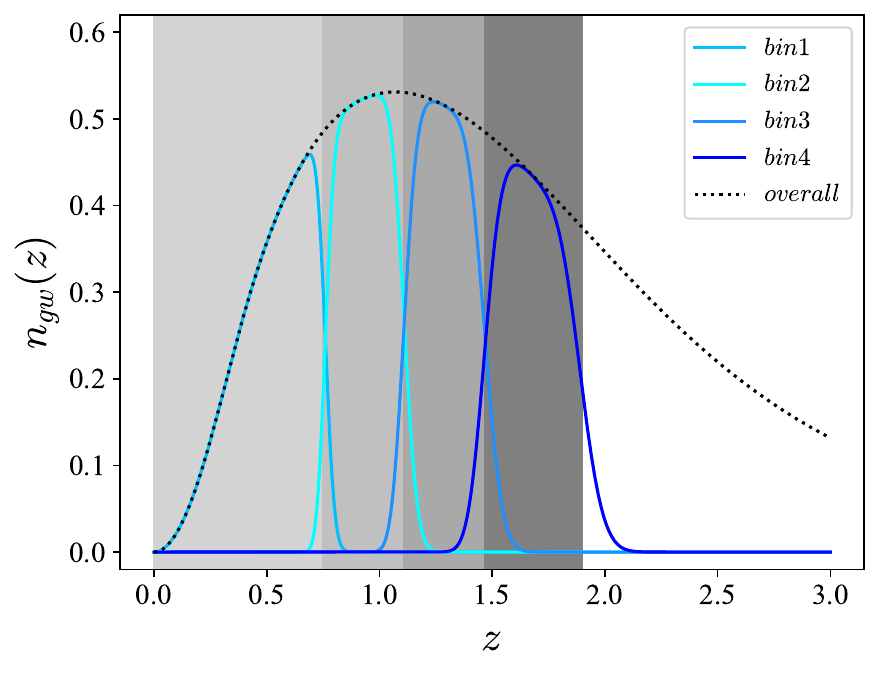}}
}
\epsscale{1.0}
    \caption{Gravitational wave sources distribution function for ECS (left panel), VKI+C (middle panel ) and HLKI+E(right panel). The shaded area corresponds to our tomographic bin.}
    \label{fig_GWdistribution}
\end{figure*}

The three detector networks exhibit distinct detection efficiencies, quantified by the detection efficiency $E_{\rm det}(z)$, representing the fraction of detectable compact binaries at redshift $z$. Previous studies generated simulated populations of BBH, BNS, and BHNS systems \citep[e.g.][]{46-borhanian2024listening}, computed signal-to-noise ratio (SNR) for each source in detector networks, and derived detection rate functions through SNR threshold filtering. Their simulations demonstrate that the detection rate curves can be well fitted by a three-parameter sigmoid function, which is given by
\begin{equation}
E_{\rm det}(z)=\left(\frac{1+b}{1+be^{az}}\right)^c.
\end{equation}
Here $a$, $b$, and $c$ are the free parameters, which are given in Table \ref{table_detec} for the three detector networks. The ECS configuration shows superior performance in both detection counts and depth, followed by VKI+C and then HLKI+E. 

In Figure~\ref{fig_DetectionRate}, we compare the detection rates for the three detector networks. Note that, for the BHNS systems, we adopt a simplified assumption $E_{\rm det}^{\rm BHNS}(z)=1/2 \left[E_{\rm det}^{\rm BBH}(z)+E_{\rm det}^{\rm BNS}(z)\right]$, due to large astrophysical uncertainties in the BHNS formation rates and the lack of standardized forecasts for the third-generation detectors. Our results are insensitive to this assumption, since the BHNS systems are expected to constitute a small fraction of the total AGWS catalog, and their contribution to the clustering signal is subdominant compared to the BNS systems. We can find that, for the detection rate, we have ECS $>$ VKI+C $>$ HLKI+E, though HLKI+E outperforms VKI+C at low redshifts ($0<z<0.5$). Considering localization capabilities, we adopt $\ell_{\rm max}=300$ for ECS and $150$ for VKI+C and HLKI+E, when computing the cross-power spectra.

\begin{table}[t]
    \centering
    \caption{ The number $N_i^{gw}$, avarage angular number density $\bar{n}_i^{gw}\;({\rm in}\ 10^{-3}\;\mathrm{arcmin}^{-2})$, and bias $b_i^{gw}$ of AGWS for the three detector networks for each tomographic bin. The effective redshift $z_i^{\rm eff}$ is computed to determine the constant bias $b_i^{gw}$ within each bin.}
    \label{table_network}
    \begin{tabular*}{\hsize}{@{}@{\extracolsep{\fill}}ccccc@{}}
    \toprule\toprule
    \multicolumn{5}{c}{\textbf{ECS}}                                    \\
    \midrule
    redshift bin & $z_i^{\rm eff}$ & $N_i^{gw}$ & $\bar{n}_i^{gw}$  & $b_i^{gw}$  \\
    \midrule
$[0.00,\; 0.90]$ & 0.65   & 13337  & 1.129           & 2.400 \\
$[0.90,\; 1.23]$ & 1.07   & 13204  & 1.117           & 2.695 \\
$[1.23,\; 1.54]$ & 1.39   & 13040  & 1.103           & 2.924 \\
$[1.54,\; 1.90]$ & 1.71   & 13133  & 1.111           & 3.170 \\
    \midrule
    \multicolumn{5}{c}{\textbf{VKI+C}}                                  \\
    \midrule
$[0.00,\; 0.83]$ & 0.59   & 5442    & 0.460            & 2.359 \\
$[0.83,\; 1.17]$ & 1.01   & 5480    & 0.464            & 2.647 \\
$[1.17,\; 1.50]$ & 1.33   & 5366    & 0.454            & 2.886 \\
$[1.50,\; 1.90]$ & 1.69   & 5364    & 0.453            & 3.152 \\
    \midrule
    \multicolumn{5}{c}{\textbf{HLKI+E}}                                 \\
    \midrule
$[0.00,\; 0.73]$ & 0.53  & 3759     & 0.318            & 2.318 \\
$[0.73,\; 1.11]$ & 0.94  & 3682     & 0.312            & 2.599 \\
$[1.11,\; 1.47]$ & 1.29  & 3740     & 0.317            & 2.853 \\
$[1.47,\; 1.90]$ & 1.68  & 3764     & 0.319            & 3.142 \\
    \bottomrule
    \end{tabular*}
\end{table}

Based on the merger rates of three compact binary types and detection efficiencies of the three detector networks, we select gravitational wave sources within the redshift range $[0,1.9]$, dividing them into 4 equal-number-count tomographic bins. For our baseline ECS configuration, each bin contains approximately $13,000$ compact binaries, while the VKI+C and HLKI+E networks have $5,400$ and $3,700$ sources per bin, respectively. When generating the mock data, we assume a constant bias $b_i^{gw}=1+1/D(z_i^{\rm eff})$ within each tomographic bin, where $z_i^{\rm eff}=\int_{z_i^{\rm min}}^{z_i^{\rm max}}z{n}^{gw}_i(z)dz$ denotes the effective redshift of the $i$-th bin, where ${n}^{gw}_i(z)=S_i(z)n^{gw}(z)$ is the GW sources distribution function for the $i$-th bin. $n^{gw}(z)=\frac{1}{\bar{n}_i^{gw}}\frac{cD_{\rm A}^2}{H(z)}\bar{n}^{gw}(z)$ is the total GW sources distribution function. We show the GW source total distribution function $n^{gw}(z)$ (black dotted curves) and distribution functions for the $i$-th bin $n^{gw}_i(z)$ (blue solid curves) in Figure \ref{fig_GWdistribution}, for the three GW dectection networks. It can be seen that the functional curve of the ECS configuration with the highest detection capability peaks at a higher redshift (left panel), and exhibits a greater peak value compared to the other two configurations (middle and right panels). The values of $z_i^{\rm eff}$, number $N_i^{gw}$, avarage angular number density $\bar{n}_i^{gw}$, and bias $b_i^{gw}$ of AGWS for the three detector networks in each bin are shown in Table \ref{table_network}.

\subsection{CSST shear measurement}
To generate CSST weak lensing mock data, we first derive the observed galaxy redshift distribution, and the Jiutian-1G simulation is utilized here, which is a high-resolution N-body simulation from the state-of-the-art Jiutian Simulation suite \citep{78-2025arXiv250321368H}. It employs an updated version of the L-Galaxies semi-analytic model to construct mock galaxy catalogs \citep{79-2015MNRAS.451.2663H,80-pei2024simulating}, incorporating CSST's instrumental design and photometric survey strategy. 

\begin{table}[t]
    \centering
    \caption{The mean redshift $\bar{z}_i$, galaxy number $N_i^g$, average angular number density $\bar{n}_i^g$ (in $\mathrm{arcmin^{-2}}$), and shear variance $\sigma^2_\gamma$ of the tomographic bins.}
    \label{tab_galaxybin}
    \begin{tabular*}{\hsize}{@{}@{\extracolsep{\fill}}ccccc@{}}
    \toprule\toprule
    redshift bin & $\bar{z}_i$ & $N_i^g$ & $\bar{n}_i^g$ & $\sigma_\gamma$ \\
    \midrule
    $[0.00,\;0.55]$ & 0.349 & $4.184\times10^8$ & 6.64 & 0.2 \\
    $[0.55,\;0.89]$ & 0.717 & $4.172\times10^8$ & 6.62 & 0.2 \\
    $[0.89,\;1.38]$ & 1.108 & $4.212\times10^8$ & 6.69 & 0.2 \\
    $[1.38,\;3.00]$ & 1.848 & $4.139\times10^8$ & 6.57 & 0.2 \\
    \bottomrule
    \end{tabular*}
\end{table}

\begin{figure}
    \centering
    \includegraphics[width=1\linewidth]{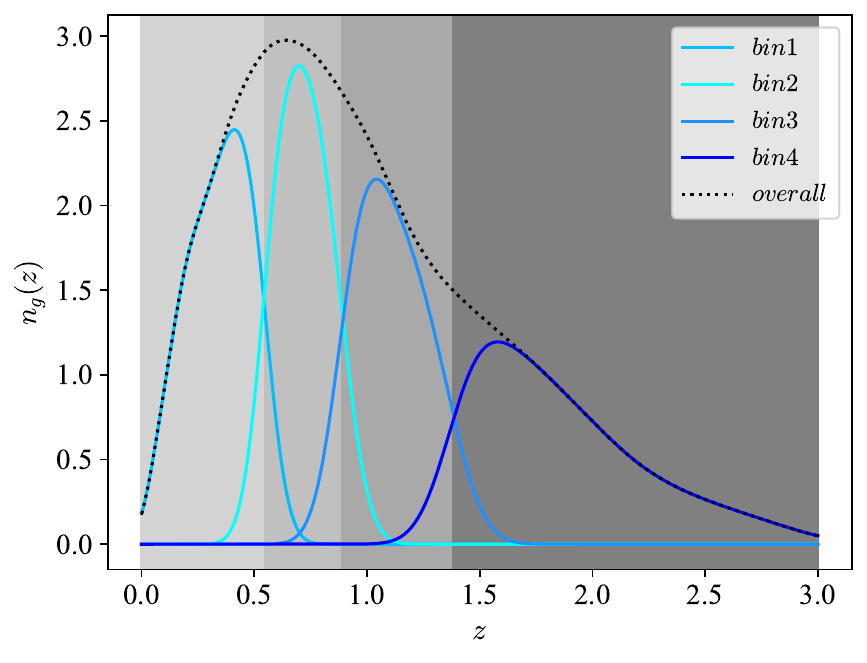}
    \caption{The galaxy redshift distribution of the CSST photometric survey derived from the simulation. The black dotted curve shows the total redshift distribution $n(z)$, and the solid curves denote the redshift distributions $n_i^g(z)$ of the four photo-$z$ bins. The shadow regions denote the redshift ranges of the four redshift bins.}
    \label{fig_Gdistribution}
\end{figure}

The photometric redshift (photo-$z$) for individual galaxies is modeled with Gaussian probability distribution function (PDF) $z_{\rm obs} \sim N(z_{\rm true},\sigma_z)$, where $\sigma_z$ is the true redshift and $\sigma_z=\sigma_{z0}(1+z)$ quantifies redshift uncertainty, assuming $\sigma_{z0}=0.05$  \citep{47-gong2019cosmology}.
While this parametrization captures the dominant contribution of photo-$z$ uncertainty, it neglects higher-order features commonly present in real photo-$z$ probability distribution, such as non-Gaussian tails, catastrophic outliers, and biases induced by template mismatch or calibration errors. These may propagate into the tomographic window functions and bias the inferred angular power spectra. In weak lensing analyses, it typically manifests as a smoothing of the cross-bin correlations and a slight degradation in cosmological parameter constraints \citep{104-2006ApJ...636...21M, 68-2006MNRAS.366..101H, 105-2021MNRAS.505.4249M}. However, in our forecast analysis, this simplification is reasonable and sufficient, and it would not affect our mian conclusions.

We divide the redshift range of the CSST photometric survey into four tomographic bins using equal-number intervals as listed in Table \ref{tab_galaxybin}. Within the CSST survey area of $17,500 \,\mathrm{deg}^2$, each bin contains approximately $4.1\times10^8$ galaxies, with a total surface density of $26.7 \; \mathrm{arcmin^{-2}}$. This result matches the prediction derived from the COSMOS catalog for the CSST photometric surveys \citep{47-gong2019cosmology}. We employ identical galaxy distributions for both lensing and source samples to better constrain the photo-$z$ uncertainty \citep[see e.g.][]{83-2020JCAP...12..001S}.

\begin{figure}
    \centering
    \includegraphics[width=1\linewidth]{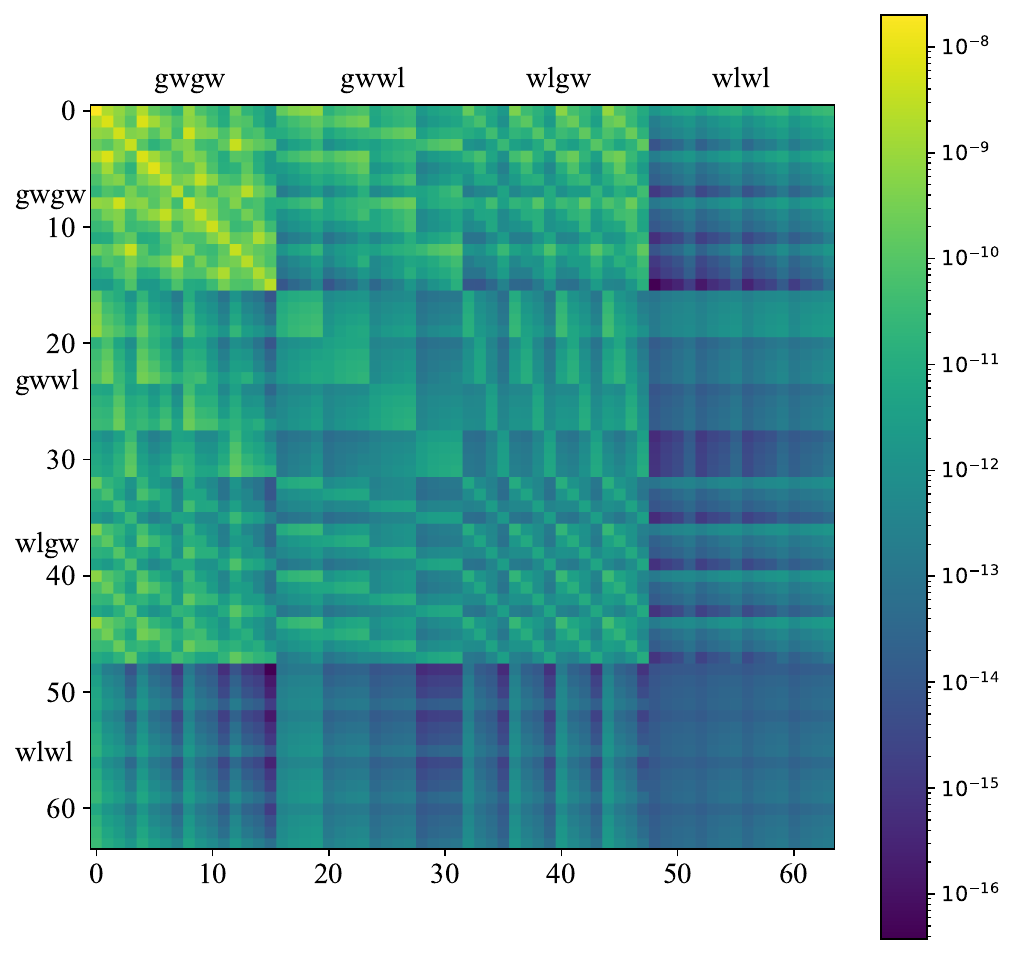}
    \caption{As an example, the covariance matrix at $\ell\simeq11$ for the joint data vector of the AGWS clustering, AGWS-weak lensing and cosmic shear power spectra for different tomographic bin combinations.}
    \label{fig_covariance}
\end{figure}

\begin{figure*}
    \centering
    \includegraphics[width=1\linewidth]{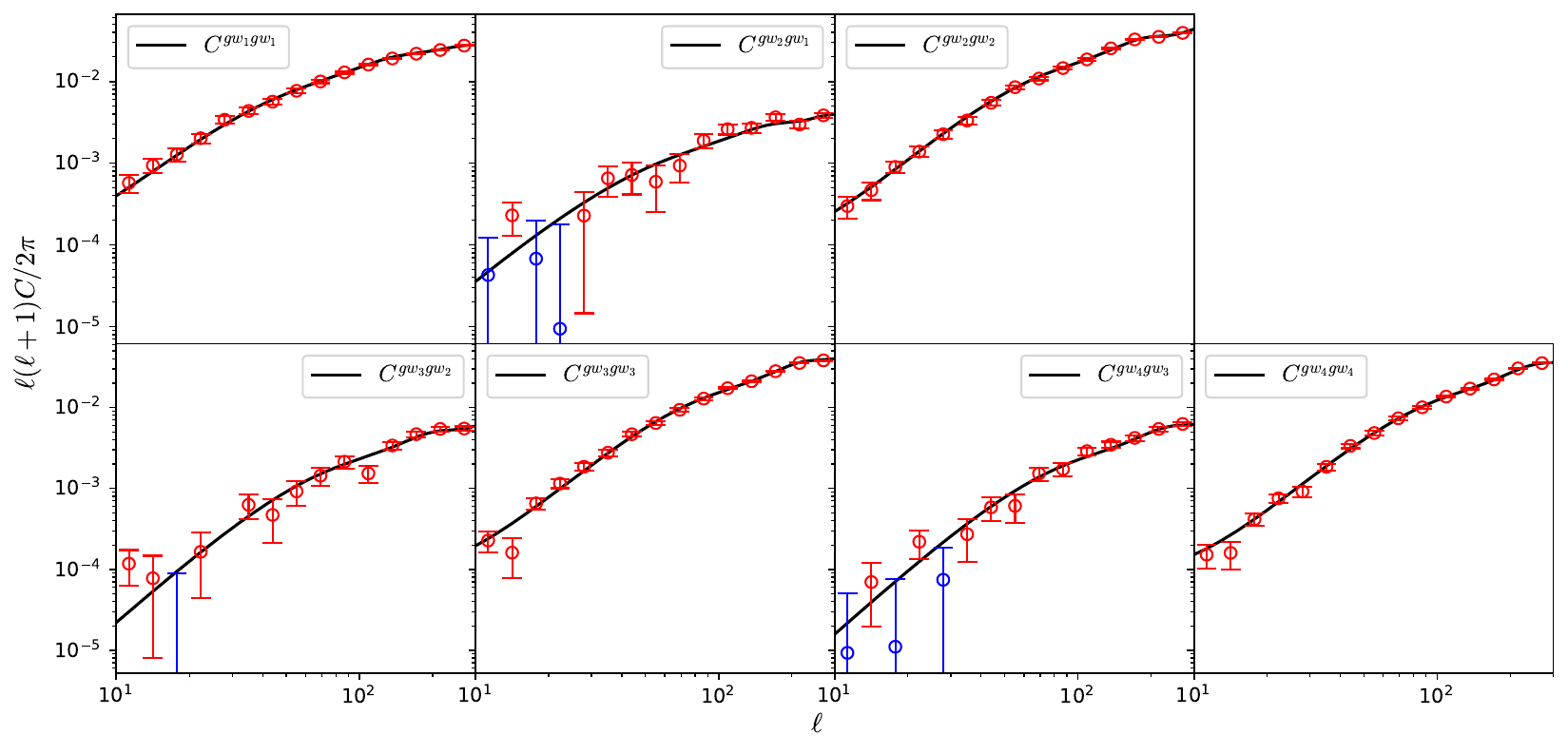}
    \caption{The mock ECS AGWS clustering auto and cross power spectra for the four tomographic bins. The solid black curves show the results of the best-fitting theoretical model. The red dots represent mock data points with $\rm SNR>1$ that will be included in the fitting process, while blue dots denote the data with $\rm SNR < 1$ which will be excluded.}
    \label{fig_GWGW}
\end{figure*}

\begin{figure*}
    \centering
    \includegraphics[width=1\linewidth]{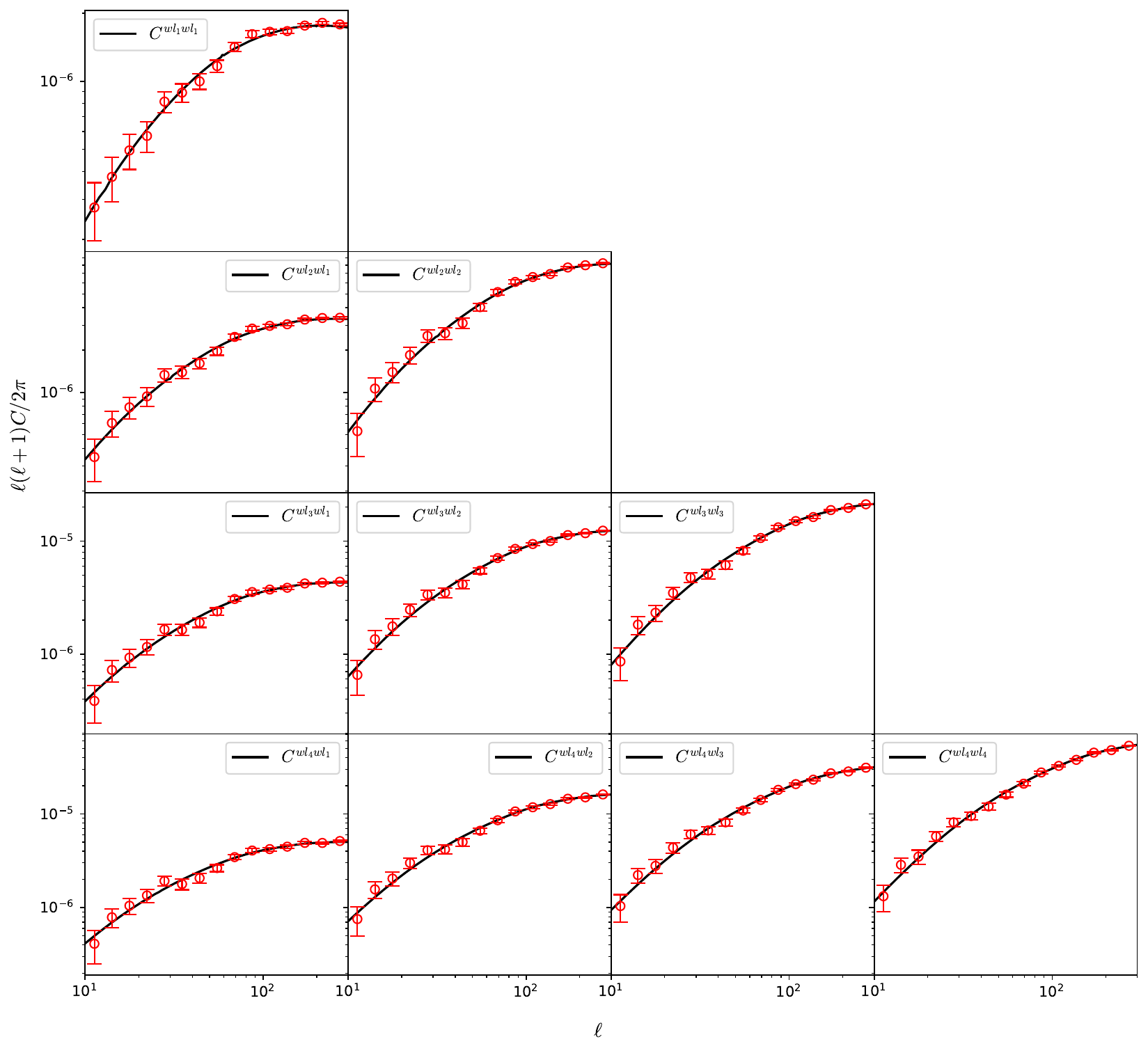}
    \caption{The mock CSST auto and cross shear power spectra for the four tomographic bins in $17500\,\mathrm{deg}^2$. The solid black curves show the best-fitting theoretical model.}
    \label{fig_WLWL}
\end{figure*}

\begin{figure*}
    \centering
    \includegraphics[width=1\linewidth]{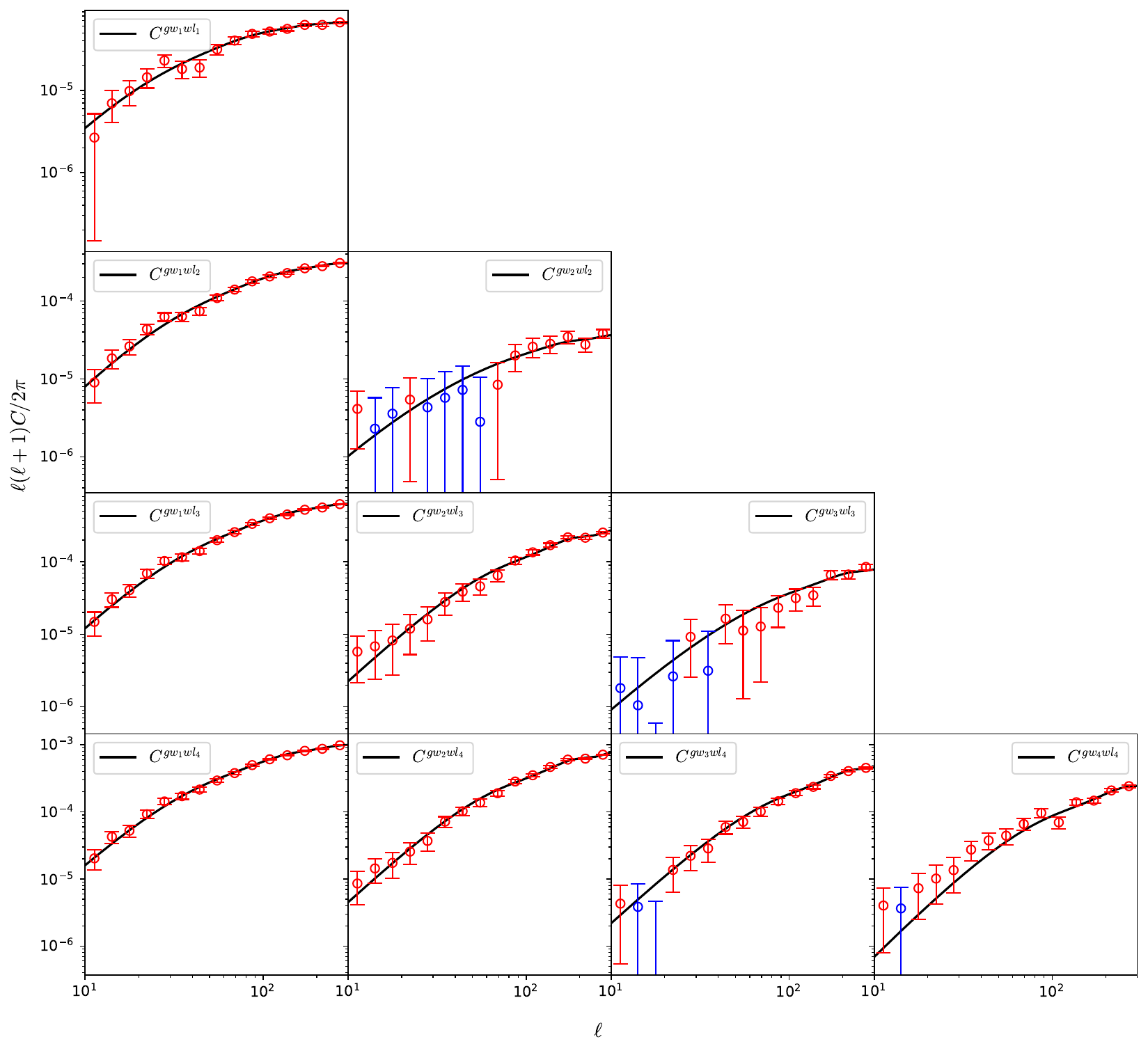}
    \caption{The mock AGWS-shear cross power spectra for the four tomographic bins in $17500\;\mathrm{deg}^2$. The solid black curves show the best-fitting theoretical model. The red dots represent mock data points with $\rm SNR>1$ that will be included in the fitting process, while blue dots denote the data with $\rm SNR < 1$ will be excluded.}
    \label{fig_GWWL}
\end{figure*}

For the normalized redshift distributions $n_i^g$ of lens and source galaxies in each tomographic bin, we introduce the shift parameter $\Delta z_i$ and stretch parameter $\sigma_i^z$ to account for potential uncertainties \citep{84-2022MNRAS.513.5517C}. 
The shift parameter $\Delta z_i$ accounts for an additive shift to the mean redshift of each bin due to the global bias in the photo-$z$ calibration, and the stretch parameter $\sigma_i^z$ denotes the uncertainty on the width of the redshift distribution \citep{85-2022PhRvD.106j3530P}.
Then the galaxy redshift distribution for each tomographic bin is given by 
\begin{align}
    n_i^g(z)\rightarrow \frac{1}{\sigma_i^z}n_i^g(\frac{(z-\bar{z}_i-\Delta z_i)}{\sigma_i^z}+\bar{z}_i), 
\end{align}
where $\bar{z}_i=\int_{z_{i,\rm min}}^{z_{i,\rm max}}zn_i^g(z)dz$ is the mean redshift of the $i$-th tomographic bin. We show the galaxy redshift distributions of the CSST photometric survey in Figure \ref{fig_Gdistribution}. More details of the galaxy catalog construction can be found in {\cite{81-2024arXiv241019388X}}. We can find that the CSST photo-$z$ distribution peaks at $z\approx0.6$, and can extend to $z>3$, which is consistent with the previous works \citep[e.g.][]{82-2018MNRAS.480.2178C,47-gong2019cosmology}.

\subsection{Covariance matrix and Mock Data}
For the AGWS clustering, cosmic shear, and the cross-power spectra, we employ a Gaussian covariance matrix to estimate the uncertainties, which can be expressed as
\begin{align}
    \mathrm{Cov}_{ijmn}^{ABCD}(\ell)&=\mathrm{Cov}[\hat{C}_{ij}^{AB}(\ell),\;\hat{C}_{mn}^{CD}(\ell')] =\frac{\delta_{\ell\ell'}}{(2\ell+1)f_{\rm sky}\Delta \ell} \nonumber \\ &\times\left(\hat{C}_{im}^{AC}(\ell)\hat{C}_{jn}^{BD}(\ell')+\hat{C}_{in}^{AD}(\ell)\hat{C}_{jm}^{BC}(\ell')\right),
\end{align}
where $f_{\rm sky}$ is the sky coverage fraction, $A,B,C,D\in \{gw,wl\}$ denote different tracers, and $i,j,m,n\in\{1,2,3,4\}$ indicate tomographic bins. We adopt $N_{\ell}=15$ logarithmic multipole bins within $\ell\in[\ell_{\rm min},\ell_{\rm max}]$, and set $\ell_{\rm max}=300$ for ECS network and $\ell_{\rm max}=150$ for VKI+C and HLKI+E networks. This choice reflects the different angular resolution and source localization capabilities of each network. Networks with longer baselines and more geographically distributed sites achieve better sky localization, enabling reliable reconstruction of small-scale (high $\ell$) clustering modes. We set $\ell_{\rm min}=10$ to ensure the validity of Limber's approximation. As an example, the computed covariance matrix at $\ell\simeq11$ is shown in Figure \ref{fig_covariance}.

The simulated data are generated through Cholesky decomposition of the covariance matrix. First, we compute a lower triangular matrix $L$ via Cholesky decomposition. A Gaussian random vector with dimension matching the simulated data ($N=N_{\ell}\times4^2$) is then generated. Multiplying this vector by $L$ and adding the result to the original angular power spectrum yields the final mock data. The mock data of the ESC AGWS clustering, CSST shear, and AGWS clustering-shear power spectra are shown in Figures \ref{fig_GWGW}, \ref{fig_WLWL}, and \ref{fig_GWWL}, respectively.

For the AGWS clustering power spectra, we retain only auto-correlations and adjacent-bin cross-correlations (i.e., 1-2, 2-3, 3-4), and discard weakly correlated bin combinations (i.e. 3-1, 4-1, 4-2) as shown in Figure \ref{WF_GW}. In the AGWS-shear cross-power spectra, we exclude $i>j$ combinations (where $i$ and $j$ denote the the tomographic bins), as the weight function $W_S^s(z)$ and $W_i^G(z)$ have a faint overlap in these bin combination (see both Figure \ref{WF_GW} and Figure \ref{WF_WL}). Besides, we only use the mock data with $\rm SNR>1$ in the fitting process.

\section{Constraint and Result} \label{Result}
\subsection{Fitting Method}
The $\chi^2$ method is adopted to fit the mock data from the GW detector networks and CSST joint observations, which takes the form as
\begin{equation}
    \chi^2=\left[\bm{D}-\bm{M}(\bm{p})\right]^T\bm{\mathrm{Cov}^{{-1}}}\left[\bm{D}-\bm{M}(\bm{p})\right]
\end{equation}
where $\bm{D}$ is the data vector of the AGWS clustering, AGWS-shear, and shear power spectra, $\bm{M}$ is the corresponding power spectra from the theoretical model, $\bm{p}$ is the model parameter vector, and $\bm{\mathrm{Cov}}$ is the covariance matrix. The likelihood function then can be calculated by $\mathcal{L}\propto \mathrm{exp}(-\chi^2/2)$. 

\begin{table}[htbp]
    \centering
    \caption{The fiducial values and flat priors of the 30 free parameters considered in our constraint process for ECS/VLI+C/HLKI+E. }
    \label{table_fiducialvalue}
    \begin{tabular*}{\hsize}{@{}@{\extracolsep{\fill}}ccc@{}}
    \toprule\toprule
    Parameter     & Fiducial Value               & Prior    \\
    \midrule
    \multicolumn{3}{c}{\textbf{Cosmology}}                           \\
    \midrule
    $\Omega_m$      & 0.3111                       &$\mathcal{U}(0.2,0.4)$          \\
    $\Omega_b$      & 0.04897                      &$\mathcal{U}(0.01,0.09)$          \\
    $h$             & 0.6766                       &$\mathcal{U}(0.55,0.85)$          \\
    $n_s$           & 0.9655                       &$\mathcal{U}(0.8,1.2)$         \\
    $w$             & -1                           &$\mathcal{U}(-1.5,-0.5)$          \\
    $\sigma_8$      & 0.8102                       &$\mathcal{U}(0.65,0.95)$          \\
    \midrule
    \multicolumn{3}{c}{\textbf{Cosmic Star Formation Rate}}                 \\
    \midrule
    $\alpha$        & 2.7                          &$\mathcal{U}(1,5)$          \\
    $\beta$         & 5.6                          &$\mathcal{U}(3,8)$          \\
    $C$             & 2.9                          &$\mathcal{U}(1,5)$          \\  
    \midrule
    \multicolumn{3}{c}{\textbf{Intrinsic Alignment}}                 \\
     \midrule
    $A_{\rm IA}$        & 1                            &$\mathcal{U}(-1,3)$          \\
    $\eta_{\rm IA}$     & 0                            &$\mathcal{U}(-1,1)$          \\
    \midrule
    \multicolumn{3}{c}{\textbf{AGWS bias (for ECS/VKI+C/HLKI+E)}}                 \\
    \midrule
    $b_1^{gw}$      & 2.400/2.359/2.318    &$\mathcal{U}(1.8,3.8)$          \\
    $b_2^{gw}$      & 2.695/2.647/2.599   &$\mathcal{U}(1.8,3.8)$          \\
    $b_3^{gw}$      & 2.924/2.886/2.853   &$\mathcal{U}(1.8,3.8)$          \\
    $b_4^{gw}$      & 3.170/3.152/3.142    &$\mathcal{U}(1.8,3.8)$          \\
    \midrule
    \multicolumn{3}{c}{\textbf{Photo-$z$ Shift \& Stretch}}           \\
    \midrule
    $\Delta z_i$    & (0,0,0,0)                    &$\mathcal{U}(-0.05,0.05)$          \\
    $\sigma_i^z$    & (1,1,1,1)                    &$\mathcal{U}(0.85,1.15)$          \\
    \midrule
    \multicolumn{3}{c}{\textbf{Shear Calibration}}                   \\
    \midrule
    $m_i$           & (0,0,0,0)                    &$\mathcal{U}(-0.1,0.1)$          \\
    \midrule
    \multicolumn{3}{c}{\textbf{Systematic noise \& Additive error}} \\
    \midrule
    $N_{\rm sys}^{gw}$    & 0                      &$\mathcal{U}(-10^{-8},10^{-8})$          \\
    $N_{\rm add}^{gwwl}$  & 0                      &$\mathcal{U}(-10^{-8},10^{-8})$          \\
    $N_{\rm add}^{wl}$    & $10^{-9}$              &$\mathcal{U}(5\times10^{-10},1.5\times10^{-9})$          \\
    \bottomrule
    \end{tabular*}
\end{table}

We make use of $\tt emcee$ to perform the fitting process \citep{86-foreman2013emcee}, which is a stable and well tested  sampler of the Markov chain Monte Carlo (MCMC). We set 112 walkers and each walker contains about 14,000 steps. We burn-in the first $2\times\mathrm{max}\{\bm{\tau}\}$ steps and thinning samples every $\frac{1}{2}\mathrm{min}\{\bm{\tau}\}$ steps for each walker, where $\bm{\tau}$ is the autocorrelation time vector. We combine all chains and obtain thousands of chain points to illustrate the PDFs of the free parameters. The fiducial values and priors of the free parameters are shown in Table~\ref{table_fiducialvalue}. We have 6 cosmological parameters, 3 cosmic star formation rate parameters,  4 gravitational wave bias parameters, as well as 17 systematical parameters in the fitting process. We employ flat prior for all parameters.

\subsection{Constraint results}
\begin{figure*}[htbp]
    \centering
    \includegraphics[width=1\linewidth]{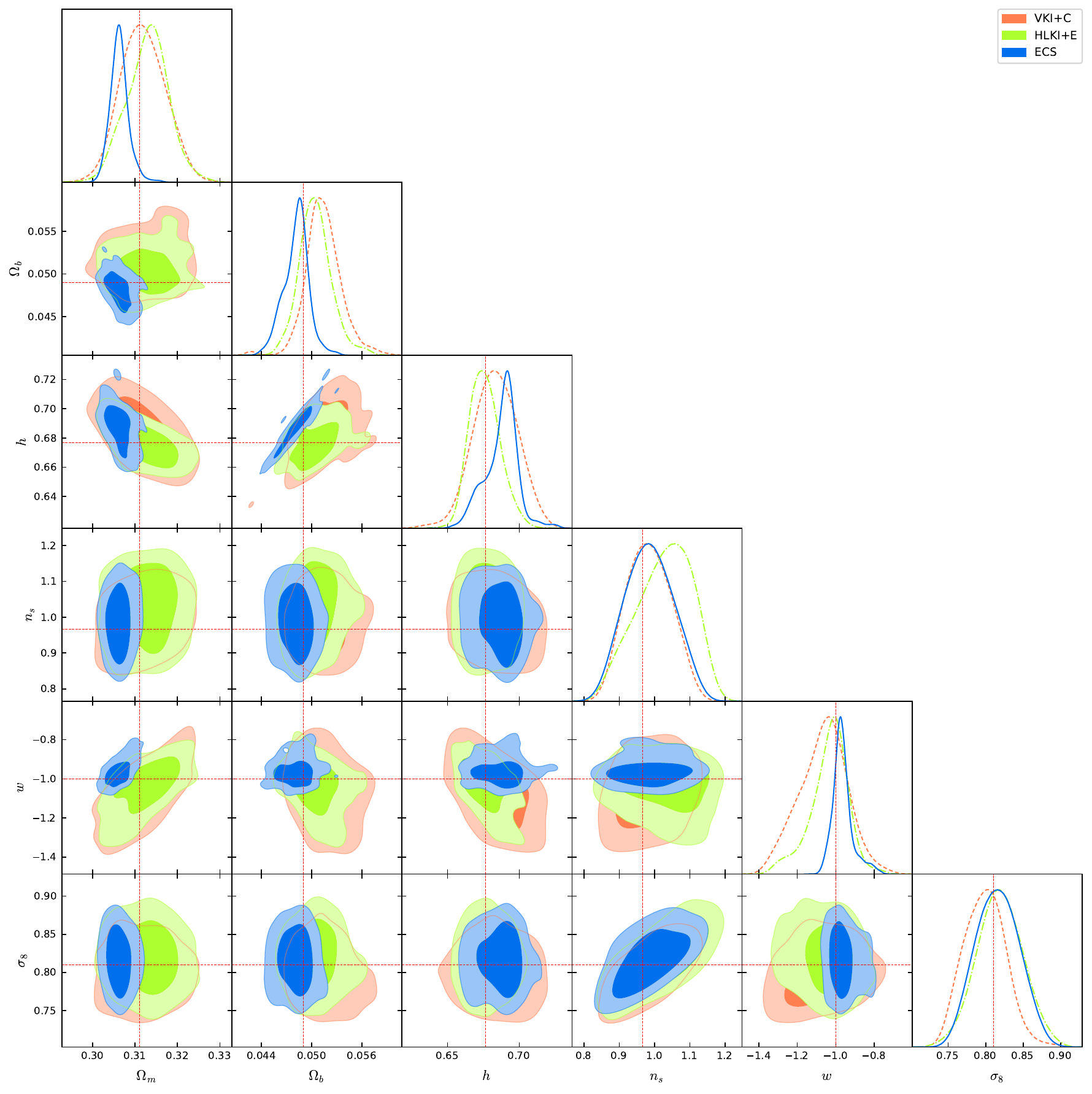}
    \caption{The contour maps of the six cosmological parameters from the joint observations of CSST with ECS (blue solid), VKI+C (red dashed), HLKI+E (green dash-dotted). The $1\sigma$ (68.3\%) and $2\sigma$ (95.5\%) CLs and the 1D PDF of each parameters are shown. The vertical and horizontal red dotted lines represent the fiducial values of these parameters.}
    \label{fig_cornerplot1}
\end{figure*}

In Figure \ref{fig_cornerplot1}, we show the marginalized contours and one-dimension (1D) PDFs for the six cosmological parameters from the joint observations of CSST with ECS, VKI+C, HLKI+E. The $1\sigma\;(68.3\%)$ and $2\sigma \; (95.5\%)$ confidence levels\;(CLs) are shown. The red dotted lines mark the fiducial values of the parameters.

\begin{table*}[htbp]
    \centering
    \caption{The joint constraint results from the CSST and the three GW detection configurations. The best-fit values, $1\sigma$ errors, and relative accuracies of the six cosmological parameters, three SFR parameters, two intrinsic alignment parameters, and four AGWS bias parameters are listed.}
    \label{table_results}
    \begin{tabular*}{\hsize}{@{}@{\extracolsep{\fill}}ccccc@{}}
    \toprule\toprule
    Parameter & Fiducial Value & \makecell{Constraints by \\ CSST \& ECS} & \makecell{Constraints by \\ CSST \& VKI+C} & \makecell{Constraints by \\ CSST \& HLKI+E} \\
    \midrule
    \multicolumn{5}{c}{\textbf{Cosmology}}                           \\
    \midrule
    $\Omega_m$      & 0.3111  &$0.306^{+0.0029}_{-0.0026}(0.90\%)$ &$0.312^{+0.008}_{-0.0073}(2.44\%)$ &0.313$^{+0.0061}_{-0.0082}(2.29\%)$         \\
    $\Omega_b$      & 0.04897 &$0.0484^{+0.0015}_{-0.0025}(4.11\%)$ &$0.0513^{+0.003}_{-0.0025}(5.4\%)$ &$0.0504^{+0.0024}_{-0.0024}(4.78\%)$          \\
    $h$             & 0.6766  &$0.691^{+0.0092}_{-0.021}(2.19\%)$ &$0.683^{+0.023}_{-0.021}(3.19\%)$ &$0.675^{+0.016}_{-0.013}(2.22\%)$          \\
    $n_s$           & 0.9655  &$0.986^{+0.11}_{-0.11}(11.20\%)$ &$0.984^{+0.1}_{-0.11}(10.49\%)$ &$1.04^{+0.11}_{-0.13}(11.64\%)$          \\
    $w$             & -1      &$-0.973^{+0.061}_{-0.05}(5.70\%)$ &$-1.06^{+0.16}_{-0.18}(16.15\%)$ &$-1.02^{+0.11}_{-0.14}(12.26\%)$          \\
    $\sigma_8$      & 0.8102  &$0.816^{+0.043}_{-0.043}(5.27\%)$ &$0.8^{+0.038}_{-0.043}(5.02\%)$ &$0.82^{+0.04}_{-0.045}(5.22\%)$          \\
    \midrule
    \multicolumn{5}{c}{\textbf{Cosmic Star formation Rate}}                 \\
    \midrule
    $\alpha$        & 2.7 &$2.9^{+0.21}_{-0.19}(6.95\%)$ &$2.94^{+0.32}_{-0.24}(9.47\%)$ &$2.81^{+0.22}_{-0.17}(6.91\%)$        \\
    $\beta$         & 5.6 &$4.98^{+0.62}_{-0.47}(10.94\%)$ &$4.77^{+1.1}_{-0.72}(18.99\%)$ &$5.68^{+1.2}_{-0.99}(19.73\%)$         \\
    $C$             & 2.9 &$2.75^{+0.19}_{-0.19}(6.99\%)$ &$2.8^{+0.36}_{-0.33}(12.31\%)$ &$3.05^{+0.28}_{-0.31}(9.64\%)$         \\
    \midrule
    \multicolumn{5}{c}{\textbf{Intrinsic Alignment}}                 \\
     \midrule
    $A_{\rm IA}$        & 1 &$0.967^{+0.057}_{-0.05}(5.53\%)$ &$0.969^{+0.059}_{-0.051}(5.67\%)$ &0.976$^{+0.049}_{-0.043}(4.76\%)$           \\
    $\eta_{\rm IA}$     & 0 &$0.183^{+0.15}_{-0.16}(86.17\%)$ &$0.219^{+0.18}_{-0.21}(88.96\%)$ &$0.19^{+0.14}_{-0.13}(72.22\%)$           \\
    \midrule
    \multicolumn{5}{c}{\textbf{AGWS bias}}                 \\
    \midrule
    $b_1^{gw}$      & (2.400,2.359,2.318)    &$2.39^{+0.089}_{-0.11}(4.16\%)$ &$2.41^{+0.12}_{-0.098}(4.48\%)$ &2.34$^{+0.11}_{-0.085}(4.17\%)$          \\
    $b_2^{gw}$      & (2.695,2.647,2.599)    &$2.69^{+0.1}_{-0.12}(4.13\%)$ &$2.71^{+0.15}_{-0.11}(4.87\%)$ &$2.63^{+0.12}_{-0.099}(4.23\%)$          \\
    $b_3^{gw}$      & (2.924,2.886,2.853)    &$2.91^{+0.11}_{-0.13}(4.08\%)$ &$2.95^{+0.17}_{-0.13}(5.06\%)$ &$2.89^{+0.15}_{-0.12}(4.62\%)$          \\
    $b_4^{gw}$      & (3.170,3.152,3.142)    &$3.19^{+0.12}_{-0.15}(4.18\%)$ &$3.29^{+0.21}_{-0.15}(5.46\%)$ &$3.24^{+0.16}_{-0.13}(4.58\%)$          \\
    \bottomrule
    \end{tabular*}
\end{table*}

\begin{figure}[htbp] 
    \centering
        \centering
        \includegraphics[width=\linewidth]{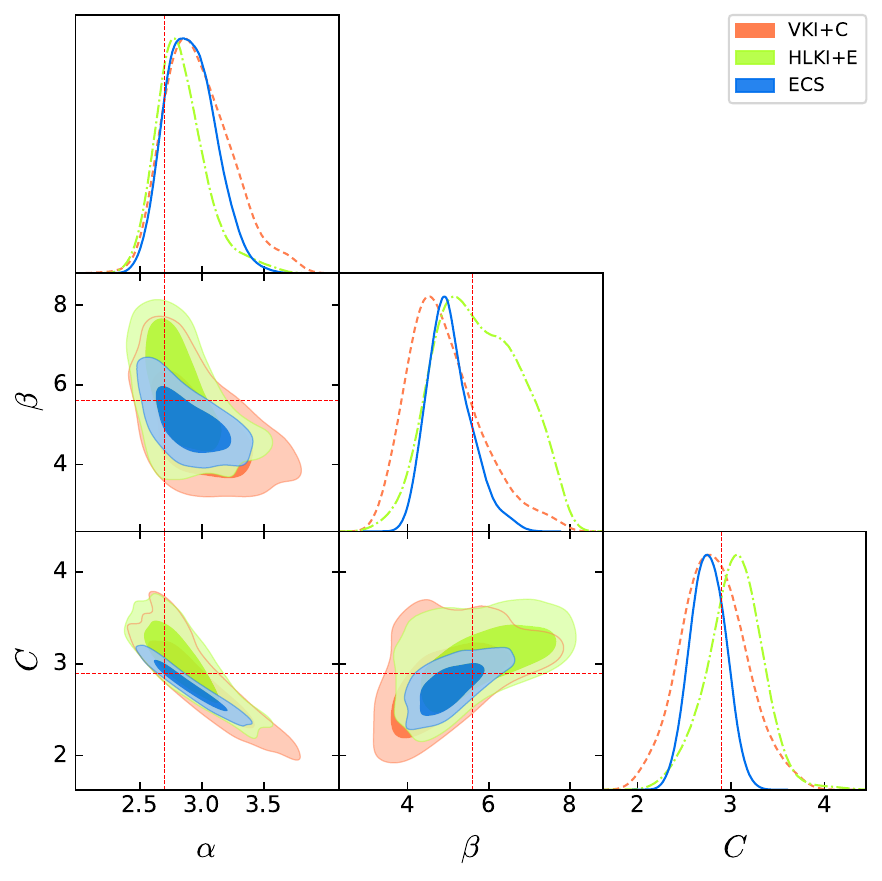}
        \caption{The contour maps of the three SFR parameters from the joint observations of CSST and ECS (blue), VKI+C (red), HLKI+E (green). The $1\sigma$ (68.3\%) and $2\sigma$ (95.5\%) CLs and the 1D PDF of each parameters are shown. The vertical and horizontal red dotted lines represent the fiducial values of these parameters.}
        \label{fig_cornerplot4}
    \end{figure}
    
    \begin{figure}[htbp] 
        \centering
        \includegraphics[width=\linewidth]{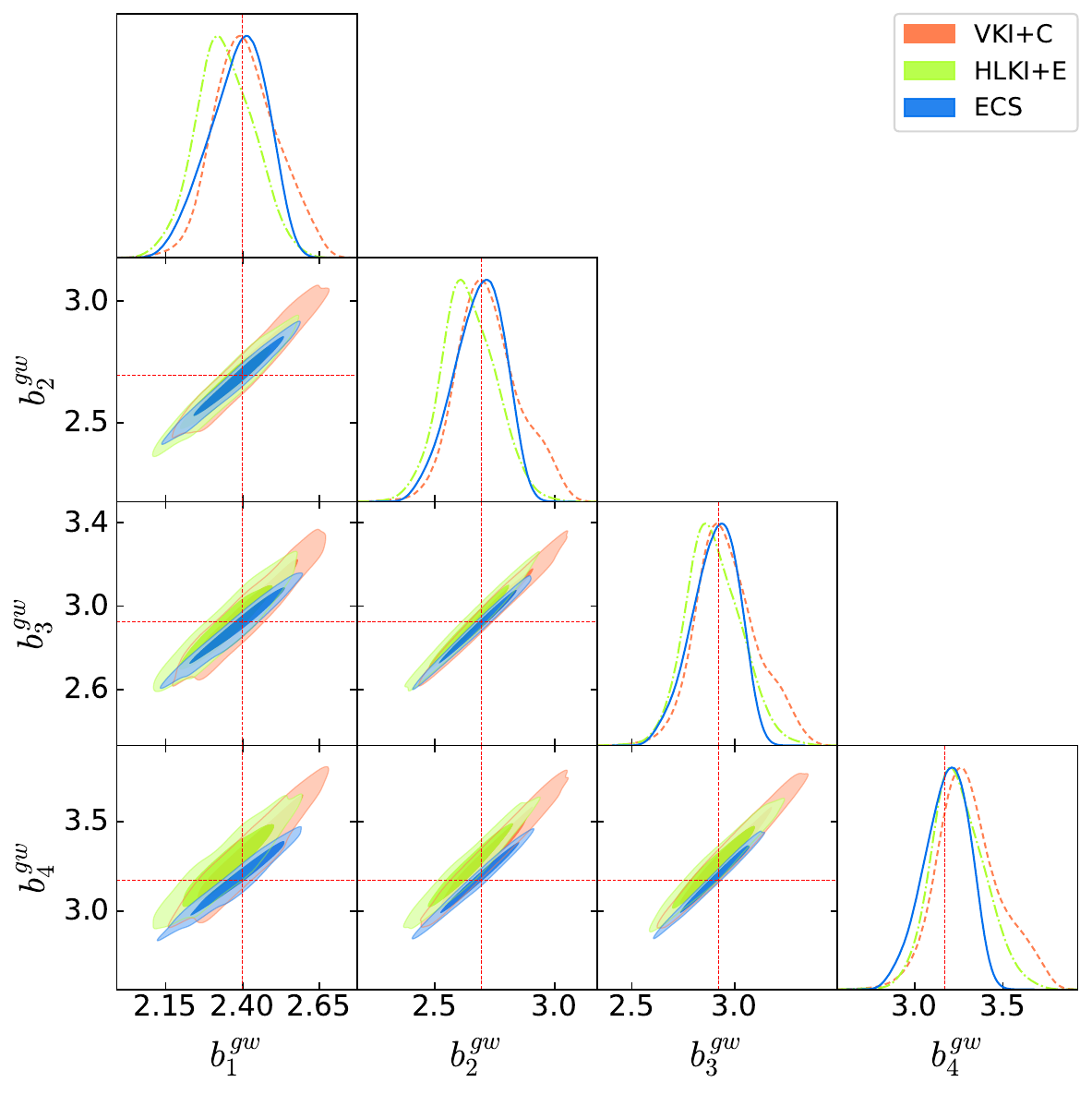}
        \caption{The contour maps of the four AGWS bias parameters from the joint observation of CSST and ECS (blue), VKI+C (red), HLKI+E (green). The $1\sigma$(68.3\%) and $2\sigma$ (95.5\%) CLs and the 1D PDF of each parameters are shown. The vertical and horizontal red dotted lines represent the fiducial values of these parameters.}
        \label{fig_cornerplot3}
\end{figure}

Our analysis demonstrates that, utilizing the mock data from CSST and ECS, the cosmological parameters can be tightly constrained. The fiducial values of most cosmological parameters lie within the $1\sigma$ CL of the fitting results. We obtain constraint accuracies of $0.9\%$, $4.1\%$, $2.2\%$, and $5.7\%$ for $\Omega_m$, $\Omega_b$, $h$, and $w$, respectively (see Table~\ref{table_results} for details), which are tighter than the other two GW detector networks. However, for $n_s$ and $\sigma_8$, we obtain constraint accuracies of $11.2\%$ and $5.3\%$, respectively, which is similar for all the three GW detector networks, since these parameters are mainly constrained by the CSST cosmic shear measurement. Besides, the results from CSST \& HLKI+E are comparable to or slightly better than that from CSST \& VKI+C, which attributed to the enhanced detection capabilities of the third-generation ground-based detectors and their coordinated multi-detector networks. 

In addition, our results indicate that future deployment of any single third-generation detector (ET or CE) could independently deliver sub-$4\%$ $H_0$ constraints, thereby providing pivotal independent constraints to resolve the Hubble tension through multi-messenger cross-validation. In terms of dark energy parameters, for the ECS configuration, we can achieve a constraint precision of $5.7\%$ on the parameter $w$, while the constraints for VKI+C and HLKI+E are less satisfactory, at $16.2\%$ and $12.3\%$ respectively.

Our constraint on the Hubble constant ($H_0=69.1^{+0.92}_{-2.1} \rm{km \;s^{-1}\;Mpc^{-1}}$ 
, i.e., $2.2\%$ precision) represents a significant improvement over current gravitational-wave-based measurements. For example, the most recent analysis from the LIGO--Virgo--KAGRA Collaboration using O4a data \citep{106-2025arXiv250904348T} yields $H_0=81.6^{+21.5}_{-15.9} \rm{km \;s^{-1}\;Mpc^{-1}}$ from the dark siren measurement alone, $H_0=78.4^{+25.7}_{-12.0} \rm{km \;s^{-1}\;Mpc^{-1}}$ from bright siren GW170817, and $H_0=76.4^{+23.0}_{-18.1} \rm{km \;s^{-1}\;Mpc^{-1}}$ from spectral siren measurement. 
Besides, compared to the cosmological constraints derived in other forecast works, e.g. using the dark siren method \citep{89-Jin_2023}, our results  are comparable to that from the future space-based gravitational wave detector networks, e.g. Taiji \citep{101-2016Natur.531..150C}, TianQin \citep{102-Luo_2016}, and Laser Interferometer Space Antenna (LISA) \citep{103-2017arXiv170200786A}.

In Figure~\ref{fig_cornerplot4}, we show the constraint results of the SFR parameters, i.e. $\alpha$, $\beta$, and $C$. As can be seen, the fiducial values of these parameters are all within the $1\sigma$ CL for the three GW detector networks, and the constraint accuracy can reach $6\%-10\%$. We can expect that, future multi-messenger observations could significantly refine the constraints and establish a novel approach for quantifying the SFR at different redshifts.

\begin{figure}[htbp]
    \centering
    \includegraphics[width=1\linewidth]{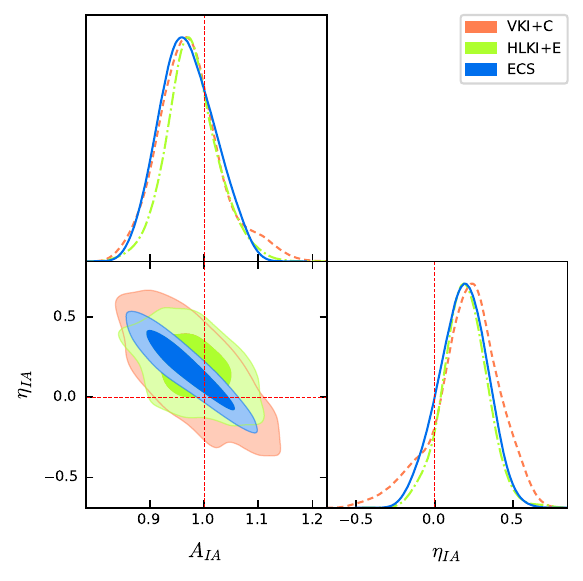}
    \caption{The contour maps of the two intrinsic alignment parameters from the joint observation of CSST with ECS\;(blue), VKI+C\;(red), HLKI+E\;(green). The $1\sigma\;(68.3\%)$ and $2\sigma \; (95.5\%)$ CLs and the 1D PDFS of each parameters are shown. The vertical and horizontal red dotted lines represent the fiducial values of these parameters.}
    \label{fig_cornerplot2}
\end{figure}

For the AGWS bias $b_i^{gw}$ in the four tomographic bins, as shown in Figure~\ref{fig_cornerplot3}, the fitting results obtained from all three detector configurations are within the $1\sigma$ CL, and the constraint accuracies are approximately $4\%-5\%$. This indicates that we can distinguish different progenitor models by measuring the AGWS bias from the angular power spectrum \citep{88-raccanelli2016determining}. This is becasue that GW events resulting from the endpoints of stellar binary evolution in a halo are expected to occur in larger and more heavily biased halos than mergers from primordial black holes (PBHs), which have been shown to occur predominantly in small halos below the threshold for forming stars.

We also find that, as shown in Figure \ref{fig_cornerplot3}, a strong positive degeneracy between the AGWS bias parameters ($b_i^{gw}$) across different redshift bins. This is because that the $\sigma_8$ parameter exhibits a negative degeneracy with the bias parameters. When the bias parameter $b_i^{gw}$ increases, the amplitude of the angular power spectrum will increases. Consequently, to maintain the original amplitude of the angular power spectrum, $\sigma_8$ must decrease. This reduction in $\sigma_8$, however, necessitates an increase in the other bias parameters to counteract the effect and preserve the amplitudes of their respective angular power spectra.

In Figure \ref{fig_cornerplot2}, the contour maps of the intrinsic alignment parameters $A_{\rm IA}$ and $\eta_{\rm IA}$ are shown for the three GW networks. We can find that there is a slight improvement on the constraint for ECS compared to other two configurations. This indicates that the joint constraint using the data from CSST and GW detector networks can be helpful to extract the redshift evaluation information of intrinsic alignment. In this work, we do not separate the intrinsic alignment signal from the total shear signal, and actually the intrinsic alignment could be an effective tool to extract the cosmological information or study galaxy dynamical evolution \citep[e.g.][]{90-Taruya_2020}. The CSST survey will have a great potential in using the intrinsic alignment as a supplementary cosmological probe, which could be studied in details in the future.

The contour maps of the redshift calibration bias $\Delta z_i$ and stretch factors $\sigma_i^z$ are shown in Figure \ref{fig_cornerplot5} in Appendix. We can find that ECS will offer a better constraint compered to VKI+C and HLKI+E, and the fiducial values of the parameters are within the $1\sigma$ CL for all three GW detector networks. The multiplication error $m_i$ and other systematical parameters (i.e. $N_{\rm sys}^{gw}$, $N_{\rm add}^{gwwl}$, and $N_{\rm add}^{wl}$) are shown in Figure \ref{fig_cornerplot6} in Appendix. In order to retain small fitting biases of the cosmological and GW source parameters (keep the fiducial values in $1 \sigma$ CL), we need to control the systematical parameters in $1\sigma$ CL at least, which requires $|\Delta z_i|<0.02$, $|\sigma_i^z|<0.05$ and $|m_i|<0.02$ for the CSST weak lensing survey \citep{47-gong2019cosmology}. Our results demonstrate that multi-messenger observational strategy can effectively suppress the systematical uncertainties, and could provide accurate constraints on the key cosmological and astrophysical parameters.

To better understand the origin of our cosmological constraints, we decompose the MCMC posterior covariance by data type and redshift bin combination. We find that the cosmological parameters sensitive to geometry and structure growth (e.g., $h$, $w$, $\Omega_m$, $\Omega_b$, $\sigma_8$, and $n_s$) are mainly constrained by the shear auto-power spectrum $C^{wlwl}(\ell)$ and cross-power spectrum $C^{gwwl}(\ell)$, which are sensitive to the matter fluctuations and expansion of the Universe. 
The AGWS auto-power spectrum $C^{gwgw}(\ell)$ primarily constrains the parameters of GW bias ($b_i^{gw}$) and the cosmic star formation rate history. When combined with $C^{gwwl}$, the degeneracies between parameters can be effectively broken and more accurate constraint results can be obtained.
Regarding the tomographic bin combinations, we find that the most informative cross-correlations are adjacent and near-diagonal bins (e.g., GW bin 2 $\times$ CSST bin 2--3, and GW bin 3 $\times$ CSST bin 3--4), which capture the peak overlap between the AGWS redshift distribution and the shear weighting function.
\section{Summary and Conclusion} \label{Conclusion}
In this work, we explore a multi-messenger framework cooperating the AGWS clustering measurements of the third generation GW detector networks and the CSST shear measurement to probe the Universe. This study presents a robust forecast for cosmological constraints by synergizing the CSST cosmic shear data with AGWS clustering from next-generation detector networks, namely ECS, VKI+C, and HLKI+E configurations. The generation of mock data involves computing theoretical angular power spectra for auto- and cross-correlations, incorporating GW selection functions and lensing systematics, then adding realistic noise via Cholesky decomposition of a Gaussian covariance matrix. By using the MCMC methods, the joint analysis achieves remarkable precision, notably constraining Hubble constant $H_0$ to $2.2\%$ and dark energy equation of state $w$ to $5.7\%$ with the ECS network, while also providing $\sim5\%$ accuracy on AGWS bias parameters. Our analysis demonstrates that the joint observation can provide stringent constraints on cosmological parameters, gravitational wave source populations, and cosmic star formation history. These findings position the next-generation GW detectors as powerful tools for the LSS cosmology and compact binary physics. 

We also notice that some assumptions made in this work can be improved to obtain more reliable and accurate results, especially for the real data analysis in the future. For instance, we can adopt more realistic or measured photo-$z$ PDFs for individual sources in the CSST weak lensing survey, instead of assuming a simplified Gaussian distribution. Besides, a more reasonable GW detection rate for the BHNS systems can be used in the future to further enhance the reliability of the results. Other effects, such as GW detector sensitivity variations across the sky and galaxy survey masks, also need to be carefully considered in the analysis of real data. Despite these challenges, the methodology developed in our work establishes a foundation for exploiting synergies between the AGWS detections and weak lensing surveys, paving the way for the LSS mapping in the multi-messenger era.

\begin{acknowledgments}
P.F.S. and Y.G. acknowledge the support from the CAS Project for Young Scientists in Basic Research (No. YSBR-092), and National Key R\&D Program of China grant Nos. 2022YFF0503404 and 2020SKA0110402. X.L.C. acknowledges the support of the National Natural Science Foundation of China through grant Nos. 11473044 and 11973047 and the Chinese Academy of Science grants ZDKYYQ20200008, QYZDJ- SSW-SLH017, XDB 23040100, and XDA15020200. This work is also supported by science research grants from the China Manned Space Project with grant Nos. CMS-CSST-2025-A02, CMS-CSST-2021-B01, and CMS-CSST-2021-A01.
\end{acknowledgments}

\begin{figure*}[htbp]
    \centering
    \includegraphics[width=0.8\linewidth]{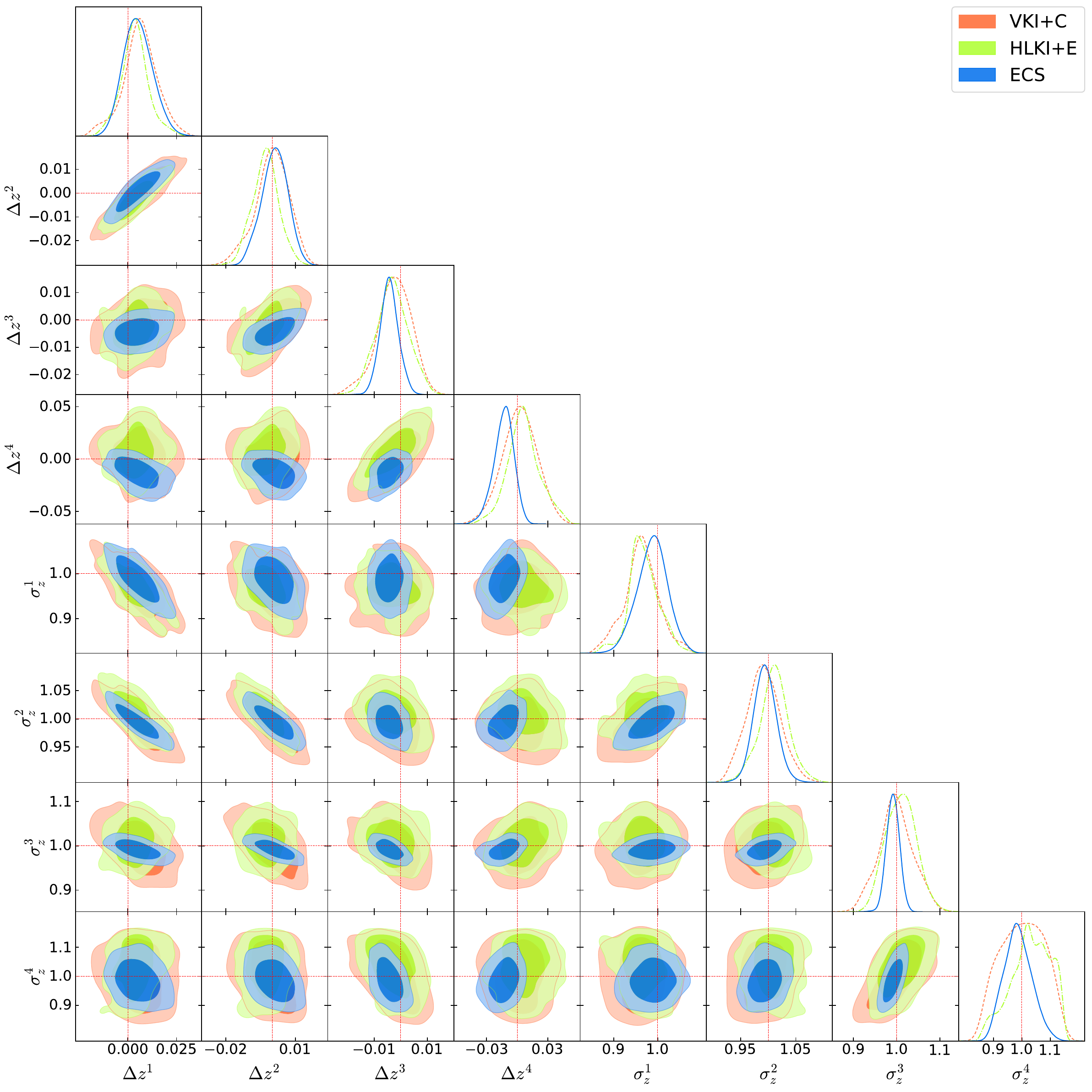}
    \caption{The contour maps of the photo-$z$ shift ($\Delta z$) and stretch ($\sigma_z$) parameters from the joint observations of CSST with ECS (blue), VKI+C (red), HLKI+E (green). The $1\sigma$ (68.3\%) and $2\sigma$ (95.5\%) CLs and the 1D PDF of each parameters are shown. The vertical and horizontal red dotted lines represent the fiducial values of these parameters.}
    \label{fig_cornerplot5}
\end{figure*}

\begin{figure*}[htbp]
    \centering
    \includegraphics[width=0.8\linewidth]{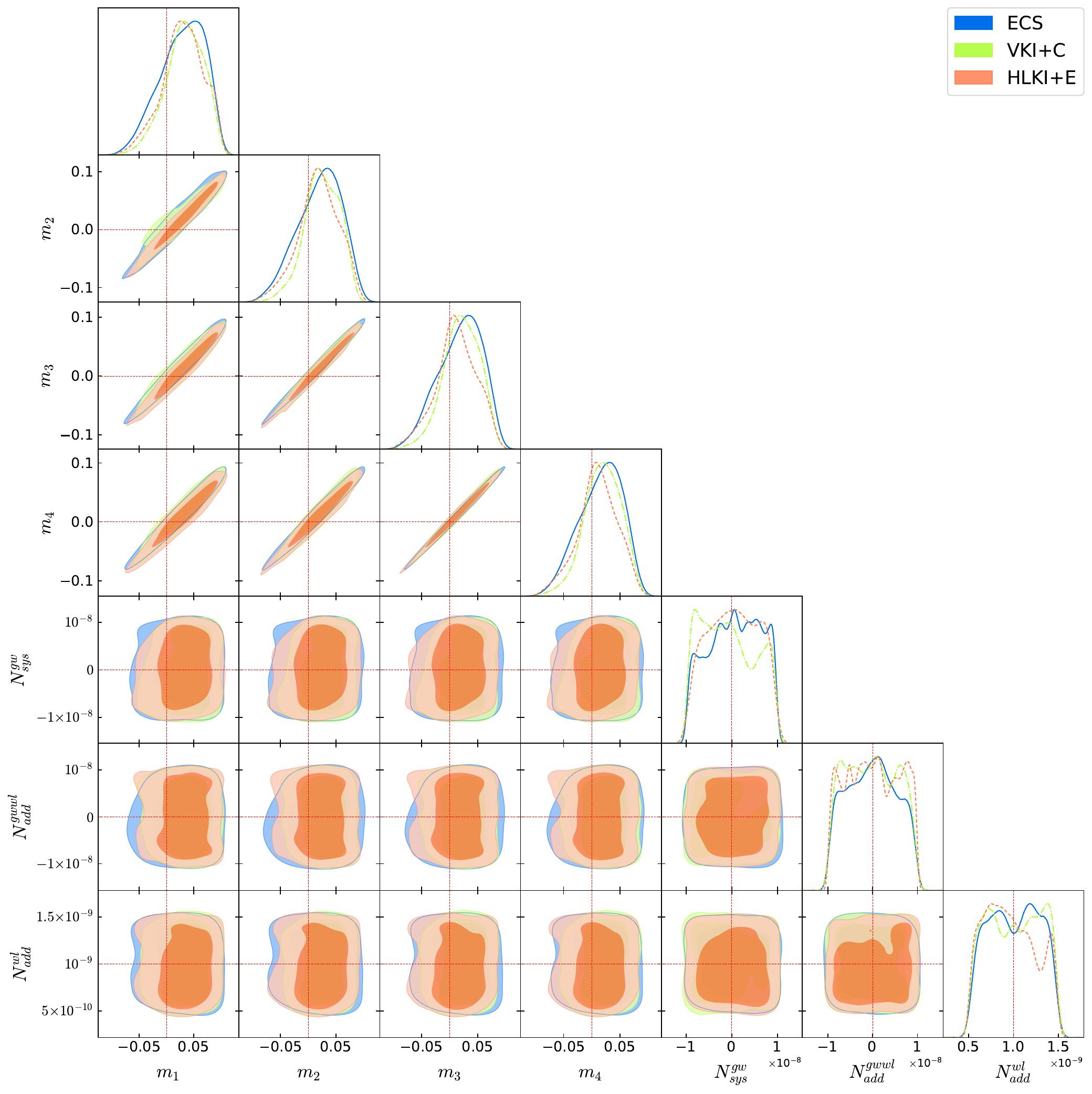}
    \caption{The contour maps of the parameters of shear calibration ($m$) and systematic noise ($N_{\rm sys}^{gw}$, $N_{\rm add}^{gwwl}$, and $N_{\rm add}^{wl}$) from the joint observations of CSST with ECS (blue), VKI+C (red), HLKI+E (green). The $1\sigma$ (68.3\%) and $2\sigma$ (95.5\%) CLs and the 1D PDF of each parameters are shown. The vertical and horizontal red dotted lines represent the fiducial values of these parameters.}
    \label{fig_cornerplot6}
\end{figure*}

\appendix
The constraint results of the photo-$z$ calibration parameters ($\Delta z$ and $\sigma_z$), the parameter of multiplicative error $m$, and the parameters of systematical noise ($N_{\rm sys}^{gw}$, $N_{\rm add}^{gwwl}$, and $N_{\rm add}^{wl}$) from the mock data of CSST and three GW detector configurations are shown in Figures~\ref{fig_cornerplot5} and \ref{fig_cornerplot6}.

\bibliography{sample7}{}
\bibliographystyle{aasjournalv7}

\end{document}